\documentclass[lettersize,journal]{IEEEtran}
\usepackage{amsmath,amsfonts}
\allowdisplaybreaks[4]
\usepackage{algorithm}
\usepackage{algorithmic}
\usepackage{array}
\usepackage{color}
\usepackage[caption=false,font=footnotesize,labelfont=rm,textfont=rm]{subfig}
\usepackage{textcomp}
\usepackage{stfloats}
\usepackage{url}
\usepackage{verbatim}
\usepackage{graphicx}
\usepackage{cite}
\usepackage{hyperref}
\usepackage[justification=centering]{caption}
\makeatletter
\newenvironment{breakablealgorithm}
  {
   \begin{center}
     \refstepcounter{algorithm}
     \hrule height.8pt depth0pt \kern2pt
     \renewcommand{\caption}[2][\relax]{
       {\raggedright\textbf{\ALG@name~\thealgorithm} ##2\par}%
       \ifx\relax##1\relax 
         \addcontentsline{loa}{algorithm}{\protect\numberline{\thealgorithm}##2}%
       \else 
         \addcontentsline{loa}{algorithm}{\protect\numberline{\thealgorithm}##1}%
       \fi
       \kern2pt\hrule\kern2pt
     }
  }{
     \kern2pt\hrule\relax
   \end{center}
  }
\makeatother
\begin{document}

\title{Non-orthogonal Multiple Access Enhanced Multi-user Semantic Communication}

\author{Weizhi Li, Haotai Liang, Chen Dong, Xiaodong Xu, ~\IEEEmembership{Senior Member,~IEEE}, Ping Zhang, ~\IEEEmembership{Fellow,~IEEE} and Kaijun Liu
\thanks{(\textit{Corresponding author: Chen Dong.})}
\thanks{This work was supported in part by the National Key Research and Development Program of China under Grant 2022YFB2902102, in part by the Fundamental Research Funds for the Central Universities (Project Number: 2482021RC01).}
\thanks{Weizhi Li, Haotai Liang, Chen Dong and Kaijun Liu are with the State Key Laboratory
of Networking and Switching Technology, Beijing University of Posts and
Telecommunications, Beijing 100876, China (e-mail: liweizhi@bupt.edu.cn; lianghaotai@bupt.edu.cn; dongchen@bupt.edu.cn; liukj@bupt.edu.cn).}
\thanks{Xiaodong Xu and Ping Zhang are with the State Key Laboratory of Networking and Switching Technology, Beijing University of Posts and Telecommunications, Beijing 100876, China, and also with the Department of Broadband Communication, Peng Cheng Laboratory, Shenzhen, Guangdong 518055, China (e-mail: xuxiaodong@bupt.edu.cn; pzhang@bupt.edu.cn).}}

\maketitle
\begin{abstract}
Semantic communication \textcolor{black}{has served} as a novel paradigm and \textcolor{black}{attracted a broad interest from researchers}. One critical aspect of it is the multi-user semantic communication theory, which can further enhance its application for the practical network environment. While most existing works have focused on the design of end-to-end single-user semantic transmission, a novel non-orthogonal multiple access (NOMA)-based multi-user semantic communication system named NOMASC is proposed in this \textcolor{black}{study}. The proposed system can support semantic transmission of multiple users with diverse modalities of source information. To avoid high demand for the hardware, an asymmetric quantizer is employed at the end of the semantic encoder for discretizing the continuous full-resolution semantic feature. In addition, a neural network model is \textcolor{black}{used} for mapping the discrete feature into self-learned symbols and accomplishing intelligent multi-user detection (MUD) at the receiver side. Simulation results demonstrate that the proposed system attains sound performance in non-orthogonal transmission of multiple user signals and outperforms the other methods, particularly with low-to-medium signal-to-noise ratios (SNRs). Moreover, it has \textcolor{black}{shown} high robustness under various simulation settings and mismatched \textcolor{black}{testing} scenarios.
\end{abstract}


\section{Introduction}
With the explosive development of information technology and emergence of numerous intelligent applications, the requirement for wireless communication technology has been proliferating. The three application \textcolor{black}{scenarios for 5G include} enhanced mobile broadband (eMBB), massive machine type communications (mMTC), and ultra-reliable and low latency communications (URLLC). In the \textcolor{black}{forthcoming} generation, the transmission rate, propagation delay, and connection density requirements \textcolor{black}{will become much greater}, which poses new challenges for wireless communication. Several candidate technologies including reconfigurable intelligent surface (RIS), terahertz communication, visible light communication (VLC), symbiotic sensing and communications (SSaC), etc. have emerged. \textcolor{black}{Nevertheless}, the traditional Shannon theory-based communication paradigm is approaching its limit under the rapidly \textcolor{black}{developing} coding and modulation schemes. To address the technical difficulty and theoretical bottleneck, semantic communication has been proposed and developed by researchers in recent years \cite{sem-com}. 

With the help of rapidly developing deep learning, the semantic communication paradigm can exploit the \textcolor{black}{data at the semantic level} \cite{weaver}. One critical technique in semantic communication is joint source-channel coding (JSCC)\cite{jscc}, which employs the autoencoder model as an end-to-end semantic codec. Unlike the Shannon paradigm, which treats each bit of information equally, semantic communication can learn the meaning and importance of the input message and saves redundant information by performing \textcolor{black}{the} source coding and channel coding coherently. \textcolor{black}{Along with} the progress \textcolor{black}{in} deep learning \textcolor{black}{from} the fields of computer vision (CV) and natural language processing (NLP), researchers \textcolor{black}{have} developed semantic communication systems for various sources, including text\cite{deepsc}, image\cite{lsci}, speech\cite{speech}, and video\cite{video}. And many remarkable \textcolor{black}{progresses have been made} for realizing highly-efficient JSCC and semantic information transmission\cite{ntscc, lsci,deepsc}. \textcolor{black}{Side information of the image has been introduced as learnable hyperprior to aid the reconstruction of image \cite{ntscc, hyperprior}. Semantic feature importance module together with the mask VQ-VAE model have been used in \cite{robust} to better counteract the semantic noise. Moreover, the resource allocation problem in semantic communication network has been solved by defining a knowledge graph and using the attention proximal
policy optimization (APPO) algorithm \cite{ARL}.}

\textcolor{black}{Current} works on semantic communication \textcolor{black}{have mostly relied} on the transmission of continuous full-resolution features, which places a heavy burden on the mobile devices with limited computing resources and makes it difficult to implement in real-world digital systems. Recently, researchers have developed digital semantic communication systems \textcolor{black}{which maps the full resolution feature into limited constellations} \cite{dtjscc,constellationjscc,jcm,jsccq,lite}. The authors \textcolor{black}{for} \cite{dtjscc} and \cite{jcm} \textcolor{black}{have employed the} Gumbel-max method\cite{gumbel} for generating discrete symbols, while in \cite{constellationjscc}, regular and irregular constellation designs \textcolor{black}{were} discussed, and a soft-to-hard quantization \textcolor{black}{was} performed by approximating the gradient \textcolor{black}{during} the back-propagation process. Tung \textit{et al}. have developed DeepJSCC-Q\cite{jsccq}, which \textcolor{black}{mapped} each element in the feature vector into the nearest symbol by generating the likelihood probability at the output of the semantic encoder. \textcolor{black}{In addition}, Xie \textit{et al}. \textcolor{black}{have also developed} a lightweight distributed semantic communication system\cite{lite}, which \textcolor{black}{used} model pruning and weight quantization tricks to reduce the model redundancy and \textcolor{black}{made} it affordable for IoT devices. 

The devices' connection density \textcolor{black}{may} also \textcolor{black}{face} a burst with more intelligent devices and users, necessitating the development of a multi-user semantic communication theory. A task-oriented multi-user semantic communication system \textcolor{black}{have been} proposed by Xie \textit{et al}. \cite{multiusercom}, which \textcolor{black}{employed} a Transformer-based model to accomplish specific intelligent tasks for \textcolor{black}{single} and multi-modal data users. Luo \textit{et al}. \textcolor{black}{have} proposed a channel-level information fusion scheme for \textcolor{black}{the} multi-user semantic communication\cite{channelwisefusion}, \textcolor{black}{which} \textcolor{black}{eliminated} the need for MUD at the receiver \textcolor{black}{side}. \textcolor{black}{Other} studies \textcolor{black}{which had taken} into account multi-user scenarios \textcolor{black}{have focused} on performing a single task like object identification using the combined messages from multiple users as \textcolor{black}{the} input \cite{objectidentification}. Additionally, semantic communication rate and spectral efficiency \textcolor{black}{were} discussed, and the resource allocation problem in the semantic-aware network \textcolor{black}{was} studied in \textcolor{black}{references} \cite{semanticrate,qoe, guocaili}.

As a promising multiple-user access scheme, NOMA\cite{noma} is capable of improving the spectrum efficiency of the system and \textcolor{black}{has been} proven to be capacity-achieving compared \textcolor{black}{with the} orthogonal multiple access (OMA) schemes. Researchers have developed several types of NOMA \cite{noma,scma,pdma}, which \textcolor{black}{may be} generally categorized into power-domain NOMA and code-domain NOMA. In NOMA transmission, signals from multiple users are overlapped on a single time or frequency resource. MUD algorithms are critical \textcolor{black}{for the design of} NOMA scheme \cite{nomamudreview}, which \textcolor{black}{can determine} its performance. Classic MUD algorithms \textcolor{black}{have included} successive interference cancellation (SIC), parallel interference cancellation (PIC), etc. In recent years, deep learning has also provided inspiration for \textcolor{black}{the design of} MUD algorithm \cite{nomamudreview,deepnoma,sicnet,pimrc}. Ye \textit{et al}. \textcolor{black}{have developed} an deep multi-task learning based framework named DeepNOMA\cite{deepnoma}. Luong \textit{et al}. \textcolor{black}{have designed} a soft decision based neural network model\cite{sicnet} for MUD, which \textcolor{black}{has attained} excellent performance even in the absence of precise channel state information (CSI). Alberge\cite{pimrc} has designed an end-to-end autoencoder model for \textcolor{black}{the} NOMA transmission, which can realize the self-design of constellation points and does not require iterative decoding. 
\begin{figure}[ht]
\centering
\includegraphics[width=3in]{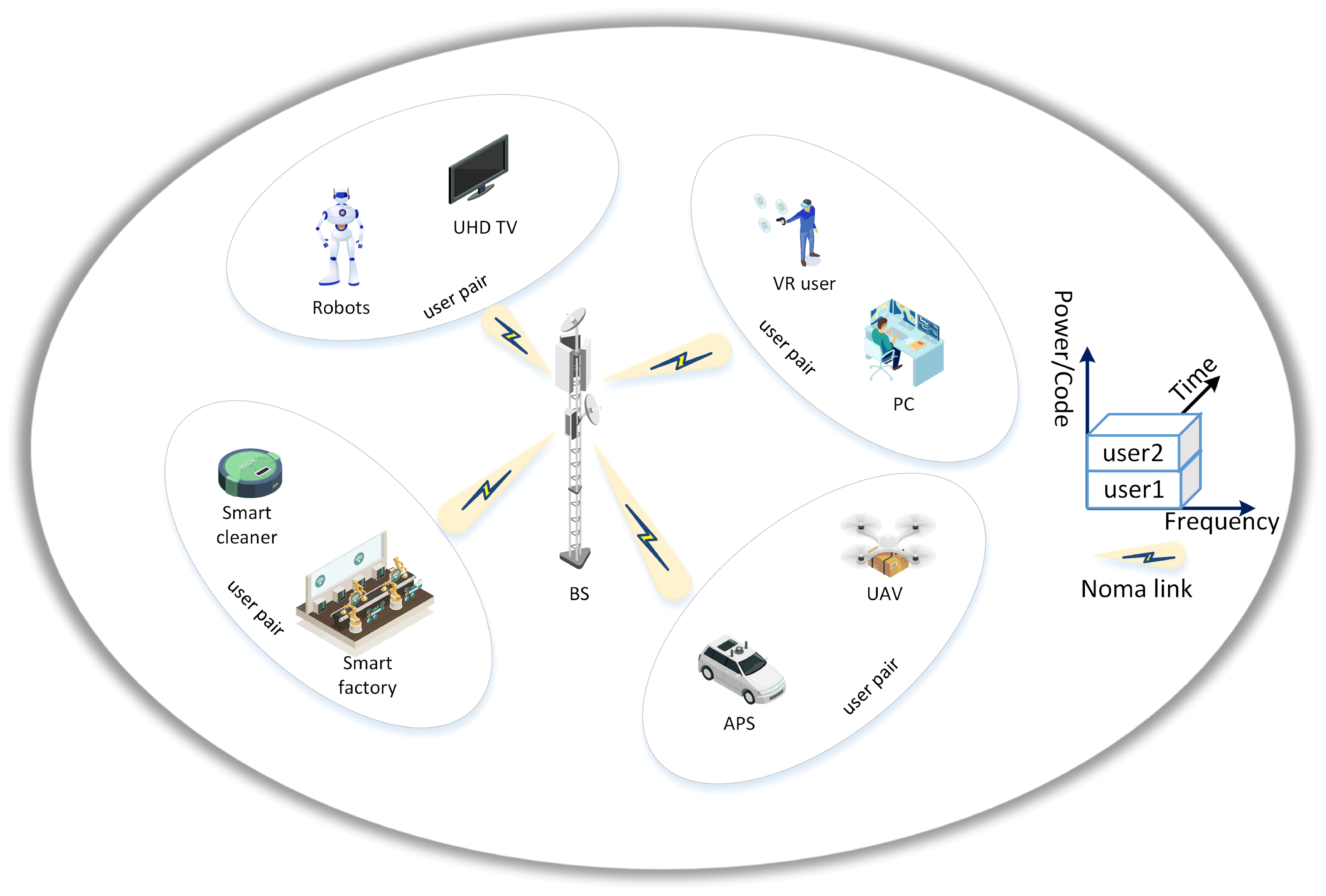}
\caption{Diagram of the proposed NOMASC system, where many intelligent semantic users are located in a cell and paired into different groups. Users within the same group are served in a non-orthogonal manner.}
\label{diagram}
\end{figure}

There are already \textcolor{black}{several} works focusing on exploiting \textcolor{black}{the} NOMA in semantic communication. Mu \textit{et al}. \textcolor{black}{have} proposed a heterogeneous semantic and bit communications framework called Semi-NOMA\cite{seminoma}, which considers transmitting \textcolor{black}{the} bit and semantic stream simultaneously. Yilmaz \textit{et al}. \cite{nomajscc} \textcolor{black}{have} introduced an image semantic compression and transmission framework which combines NOMA and DeepJSCC. However, some limitations in the existing NOMA-based semantic communication system still hinder its \textcolor{black}{use in practice}, which will be covered in section \uppercase\expandafter{\romannumeral4}. \textcolor{black}{Different from the previous works, only the transmission of pure semantic signal is considered in this paper. We \textcolor{black}{focus} on creating a general paradigm for realizing non-orthogonal semantic transmission for users with arbitrary datasets and various data modalities. In addition, to make better interconnection with the existing digital communication framework and protocols, we take modulation on the encoded semantic signal, which has been ignored by the majority of previous works. The processes of demodulation and MUD are integrated naturally using a single lightweight neural network, eliminating the needs for tedious operations. Not only did we consider the practical end-to-end transmission scheme, but also the theoretical perspective of the proposed system to give a thorough demonstration.} The diagram of the proposed system is shown in Fig. \ref{diagram}. The aim of our design is to enable multi-user semantic communication with higher transmission rate and efficiency using learning-based approaches. The following is a summary of the contributions made by this paper:
\begin{itemize}
  \item [1)]
  A NOMA-based multi-user semantic communication system named NOMASC is proposed, \textcolor{black}{which is} capable of serving user pairs with \textcolor{black}{various} types of datasets and different modalities of data. The semantic symbols of all users in the same group are superimposed and transmitted in the downlink channel using the same time or frequency resource, thereby enhancing \textcolor{black}{the} spectral efficiency.
  \item [2)]
  The theoretical rate region and power region of the proposed NOMASC scheme are characterized using the concept of semantic transmission rate and compared \textcolor{black}{with} the OMA scenario to verify the advantage of the proposed scheme.
  \item [3)]
  An asymmetric quantizer is adopted for mapping the full-resolution semantic feature vector to discrete constellations. A neural network based modem model is proposed to accomplish the modulation and demodulation processes. The modulation constellation point set is generated adaptively to the channel condition and power allocation, which can be regarded as an unequal protection for the semantic information.
  \item [4)]
  Extensive experiments have been conducted to verify the performance of the proposed system. And our results show that the proposed model (a) \textcolor{black}{has outperformed} various baseline methods, including traditional separate source and channel coding and a number of learning-based methods under both AWGN and Rayleigh channels, (b) retain good performance and high robustness under \textcolor{black}{various testing} environments and settings, and (c) make flexible grouping possible for users with various data needs.
\end{itemize}
The organization of this paper is as follows: Section \uppercase\expandafter{\romannumeral2} \textcolor{black}{will introduce} the general framework, detailed design, and training procedure \textcolor{black}{for} the proposed NOMASC. Section \uppercase\expandafter{\romannumeral3} gives the theoretical analysis of rate region and power region. Simulations settings, numerical results, and complexity analysis are presented in section \uppercase\expandafter{\romannumeral4}. Conclusions are drawn in section \uppercase\expandafter{\romannumeral5}.

\section{NOMA-based semantic communication}
\subsection{General Framework}
\begin{figure*}[ht]
\centering
\includegraphics[width=5in]{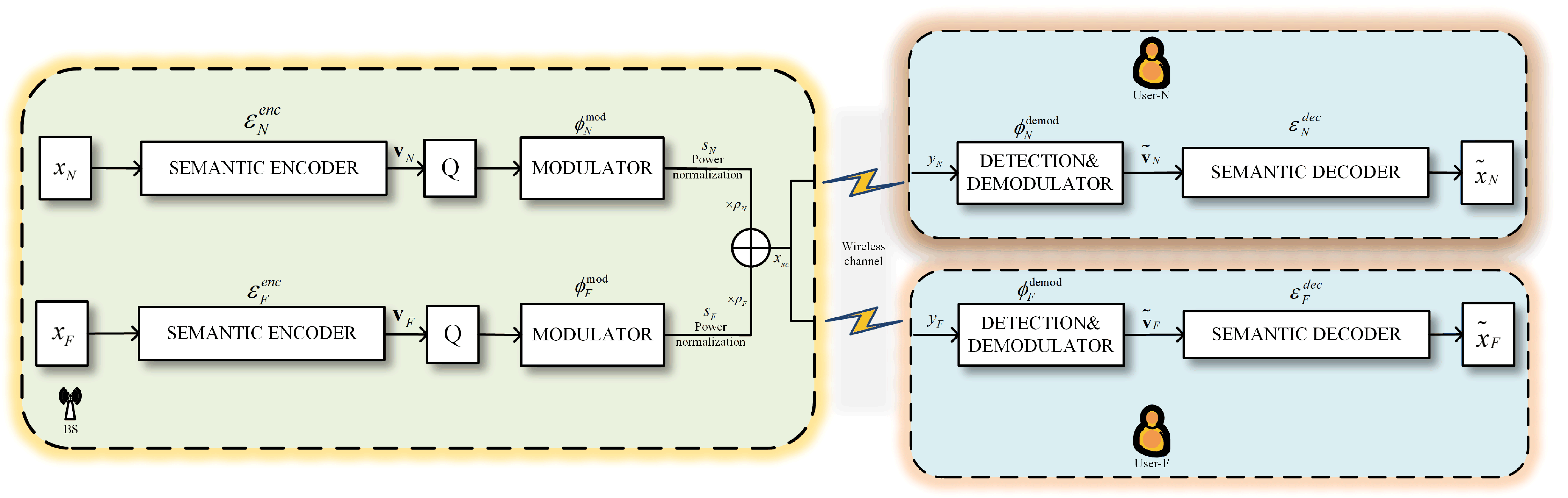}
\caption{The framework of \textcolor{black}{the} NOMASC. The semantic encoder and decoder are \textcolor{black}{used for} compressing and reconstructing \textcolor{black}{the} semantic information, the quantizer is used for mapping the continuous feature vector into limited constellations, and the modulator and demodulator can map the quantized vector into two-dimensional symbols and carry out the detection \textcolor{black}{for} the user signals.}
\label{generalframework}
\end{figure*}

The general paradigm of the proposed NOMASC system is depicted in Fig. \ref{generalframework}. Considering the downlink NOMA transmission scenario, where all users in a single cell are grouped according to their channel conditions. Each group is assumed to have two users, named user-N (near user) and user-F (far user). User-N has a better downlink transmission channel condition than user-F. Both user-N and user-F are semantic users, which means that they \textcolor{black}{will} require semantic information from base station (BS). The categories and modalities of the information required by \textcolor{black}{the} two users can be different, e.g., \textcolor{black}{whilst} user-N may require high-resolution medical images, user-F \textcolor{black}{may require} photos of cute animals. During \textcolor{black}{the} downlink transmission, the original data $\mathbf{x}_{N}, \mathbf{x}_{F}$ required by user-N and user-F are first compressed by the semantic encoder into full resolution vectors $\mathbf{v}_{N}=\boldsymbol{\varepsilon}_{N}^{enc}\left(\mathbf{x}_{N}\right), \mathbf{v}_{F}=\boldsymbol{\varepsilon}_{F}^{enc}\left(\mathbf{x}_{F}\right)$. \textcolor{black}{Subsequently}, an asymmetric quantizer is employed for mapping the feature vectors $\mathbf{v}_{N}$ and $\mathbf{v}_{F}$ into limited constellations. The quantization and modulation processes can be denoted by:
\begin{equation}
\label{deqn_ex1a}
\mathbf{s}_{i}=\boldsymbol{\phi}_{i}^{mod}\left(\mathcal{Q}\left(\mathbf{v}_{i}\right)\right), \ i\in\{N,F\}.
\end{equation}
In order to satisfy the power constraint, $\mathbf{s}_{N}$ and $\mathbf{s}_{F}$ are normalized as:
\begin{equation}
\label{deqn_ex1a}
\textcolor{black}{\mathbf{s}_{i}^{norm}=\mathbf{s}_{i} / \sqrt{\mathbb{E}\left[|\mathbf{s}_{i}|^{2}\right]}, \ i\in\{N,F\}},
\end{equation}
where the maximum transmission power of BS is $P_{max}$. \textcolor{black}{Similar with the} traditional NOMA, signal $\mathbf{s}_{N}^{norm}$ and $\mathbf{s}_{F}^{norm}$ are superimpose-coded at BS as follows:
\begin{equation}
\label{deqn_ex1a}
\mathbf{x}_{sc}=\rho_{N}P_{max}\mathbf{s}_{N}^{norm}+\rho_{F}P_{max}\mathbf{s}_{F}^{norm},
\end{equation}
where $\rho_{N}$ and $\rho_{F}$ are the power allocation factors for user-N and user-F \textcolor{black}{which satisfies} $\rho_{N}+\rho_{F}=1$. \textcolor{black}{In the context of semantic communication, power allocation algorithms such as full search power allocation (FSPA), fractional power allocation (FTPA), and fixed power allocation (FPA) may be used \cite{power_allo}. These algorithms take the channel gains of different users as inputs and use the weighted sum of semantic rates or user fairness as optimization criteria to calculate the power allocation factor for each user.} The superimposed signal $\mathbf{x}_{sc}$ is propagated through the wireless channel. And the received signals of user-N and user-F can be denoted by:
\begin{equation}
\label{deqn_ex1a}
\mathbf{y}_{i} = \mathbf{h}_{i}\mathbf{x}_{sc}+\mathbf{n}_{i}, \ i\in\{N,F\},
\end{equation}
where $\mathbf{n}_{N}\sim\mathcal{CN}(0, \sigma_{N}^{2})$ and $\mathbf{n}_{F}\sim\mathcal{CN}(0, \sigma_{F}^{2})$ is the additive white Gaussian noise (AWGN) at the receiver following a zero-mean complex normal distribution. After equalization, the estimated symbols $\widetilde{\mathbf{s}_{N}}$ and $\widetilde{\mathbf{s}_{F}}$ can be \textcolor{black}{derived with the} MUD algorithms, and the desired data $\widetilde{\mathbf{x}_{N}}$ and $\widetilde{\mathbf{x}_{F}}$ is reconstructed through \textcolor{black}{a} locally deployed demodulator and semantic decoder, which are indicated as:
\begin{equation}
\label{deqn_ex1a}
\widetilde{\mathbf{x}_{i}} = \boldsymbol{\varepsilon}^{dec}_{i}(\boldsymbol{\phi}_{i}^{demod}(\widetilde{\mathbf{s}_{i}})), \ i\in\{N,F\}.
\end{equation}

\subsection{Semantic Encoder and Decoder}
In the proposed system, users with various types and modalities of sources are considered to be served by \textcolor{black}{the} BS. Therefore, \textcolor{black}{various} semantic codecs have to be designed and deployed on \textcolor{black}{the} users' devices. We use the GAN-based semantic image codec (LSCI)\cite{lsci} for processing image-type sources\textcolor{black}{. The} base model \textcolor{black}{for} LSCI can be denoted as $\boldsymbol{\varepsilon}^{I}(\boldsymbol{\theta}_{\boldsymbol{\varepsilon}}^{I})$, with $\boldsymbol{\theta}_{\boldsymbol{\varepsilon}}^{I}$ representing the set of trainable parameters in the model, \textcolor{black}{which} is composed of three parts: the encoder $\boldsymbol{\varepsilon}^{I,enc}$, generator $\boldsymbol{\varepsilon}^{I, dec}$ and the discriminator $\boldsymbol{\varepsilon}^{I, dis}$. During the forward propagation, the input images \textcolor{black}{will} go through a series of convolution layers and output a latent feature vector. The number of output channels in the last convolution layer controls the image compression ratio (Cr). After going through a number of operations and being transmitted over the wireless channel, the feature matrix is fed into the generator for reconstructing the original image. The discriminator $\boldsymbol{\varepsilon}^{I, dis}$ is used only for training and seeks to distinguish the original and reconstructed image.

As for \textcolor{black}{the} processing \textcolor{black}{of} textual sources, the Transformer-based text semantic codec (DeepSC)\cite{deepsc} is employed. It mainly contains a pair of semantic codec $\boldsymbol{\varepsilon}^{S,enc}$ and $\boldsymbol{\varepsilon}^{S,dec}$, a pair of channel codec $\boldsymbol{\varepsilon}^{S,chenc}$ and $\boldsymbol{\varepsilon}^{S,chdec}$, and a dense layer model $\boldsymbol{\varepsilon}^{S,FC}$, with $\boldsymbol{\theta}_{\boldsymbol{\varepsilon}}^{S}$ being the trainable parameter set of the whole model. $\boldsymbol{\varepsilon}^{S,enc}$ and $\boldsymbol{\varepsilon}^{S,dec}$ are capable of completing end-to-end sentence compression and reconstruction using Transformer blocks and attention mechanism. The channel codec $\boldsymbol{\varepsilon}^{S,chenc}$ and $\boldsymbol{\varepsilon}^{S,chdec}$ are responsible for adding redundancy to the encoder output and shielding it against \textcolor{black}{the} noise and interference.

The detailed design of the LSCI and DeepSC model structures can be found in \cite{lsci} and \cite{deepsc}, which will be omitted in this paper. 

\subsection{Neural Modulator, Demodulator, and Detection} 
\begin{figure}[ht]
\centering
\includegraphics[width=3in]{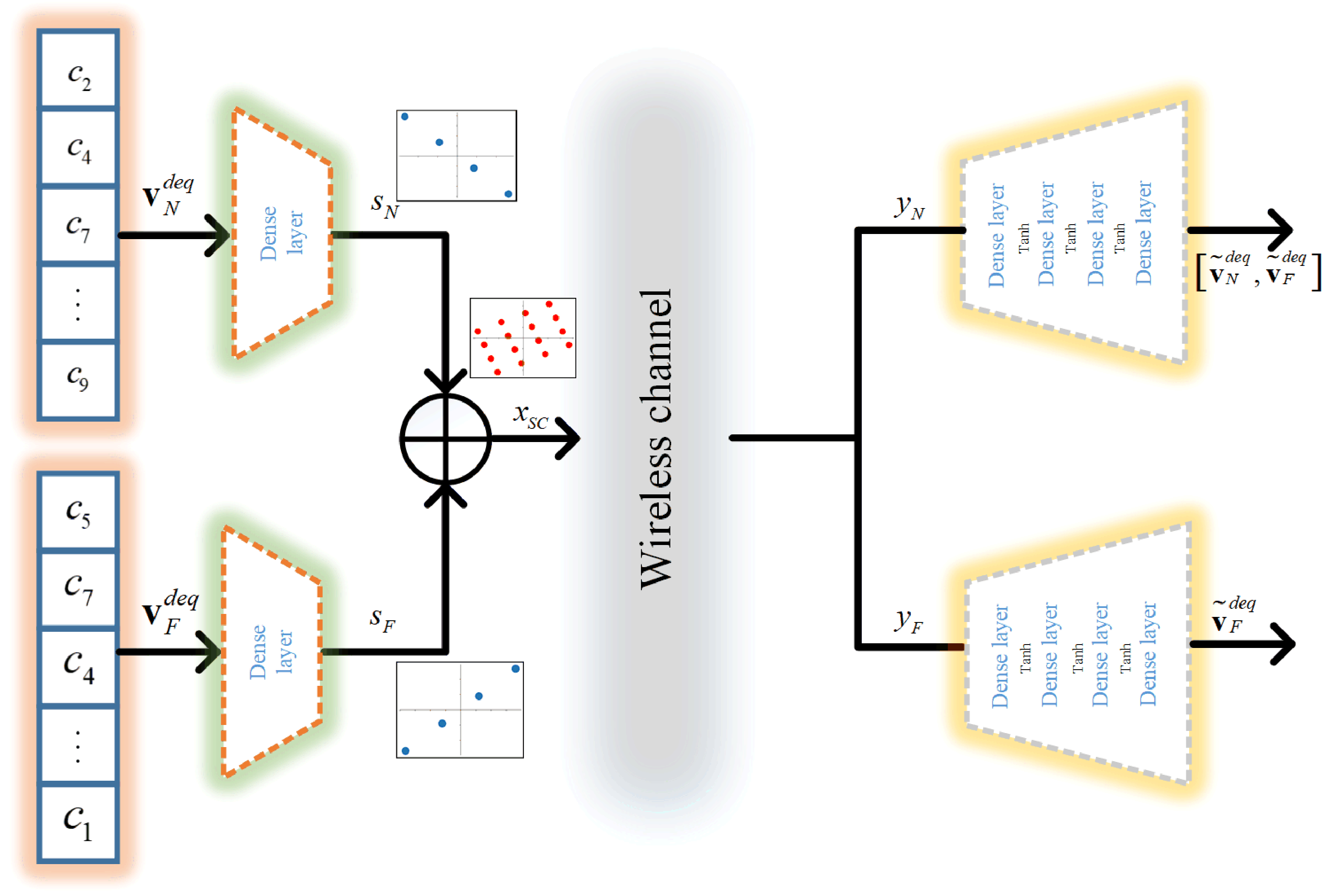}
\caption{The structure of the modem model, which \textcolor{black}{is comprised of} dense layers, is capable of \textcolor{black}{completing} the mapping between quantized vectors and two-dimensional symbols as well as user signal detection.}
\label{modulationmodel}
\end{figure}

\textcolor{black}{To detect} each user's signal at the receiver \textcolor{black}{side} is critical to the \textcolor{black}{design of the system}. Traditional MUD techniques like SIC detects user signals iteratively. For instance, in two-user scenarios, the signal of a weak user can be decoded directly by treating the signal of the strong user as noise. For the strong user, \textcolor{black}{the signal} must first be decoded and subtracted from the received signal before the strong user can decode the signal itself. However, this may lead to \textcolor{black}{greater} complexity and time consumption. \textcolor{black}{In this study, we have proposed} a neural network based end-to-end lightweight auto-encoder for modulation, demodulation, and signal detection. The structure of the modem model is shown in Fig. \ref{modulationmodel}.

Similar to the design in \cite{pimrc}. the modulator models $\boldsymbol{\phi}^{mod}$ at user-N and user-F both \textcolor{black}{comprise a} dense layer to complete the mapping from the one-dimensional quantization constellation to the two-dimensional symbol, which corresponds to the real and imaginary parts of the symbols. Identity mapping is used as an activation function for the dense layer. Therefore, the modulator \textcolor{black}{may} be regarded as a linear transform. The position and distance between each constellation symbol are optimized during \textcolor{black}{the} training, which eliminates the need for manual design. The decoder $\boldsymbol{\phi}^{demod}$ serves as a maximum likelihood estimator of $\mathbf{v}^{deq}$, which \textcolor{black}{comprises} four dense layers. User-F only detects the signal itself. For user-N, $\mathbf{v}_{N}^{deq}$ and $\mathbf{v}_{F}^{deq}$ are detected simultaneously, since $\mathbf{v}_{F}^{deq}$ is the major component in the superimposed signal $\mathbf{x}_{sc}$. The trainable parameter set of the entire modem model is represented by $\boldsymbol{\theta}_{\boldsymbol{\phi}}$.
\subsection{Asymmetric Quantizer} 
An asymmetric quantizer is employed to discretize the continuous output of the semantic encoder. For a $m$-bit quantizer, the float number in the feature vector $\mathbf{v}$ is mapped to a finite constellation set with $2^{m}$ integers. The scale factor and zero point for quantization can be calculated as follows:
\begin{equation}
\label{deqn_ex1a}
f_{s} = \frac{2^{m}-1}{max(\mathbf{v})-min(\mathbf{v})},
\end{equation}
\begin{equation}
\label{deqn_ex1a}
p_{z} = round(min(\mathbf{v})\times f_{s}).
\end{equation}
For each element $x_{i}$ in $\mathbf{v}$, the quantization process can be expressed as:
\begin{equation}
\label{xq}
x_{q} = clamp\left(round\left(x_{i}*f_{s}-p_{z}\right), 0, \ 2^{m} - 1\right),
\end{equation}
where the clamp operation is:
\begin{equation}
\label{clamp}
clamp(x_{i}, 0, 2^{m}-1) = \begin{cases}
    0, \ x_{i} < 0\\
    x, \ 0\leq x_{i}\leq2^{m}-1\\
    2^{m}-1, \ x_{i}>2^{m}-1.
\end{cases}
\end{equation}
After quantization, each element $x_{q}$ \textcolor{black}{will take} a value from the finite integer set. However, the loss of precision may cause a severe degradation of \textcolor{black}{the} model performance. Dequantization operation is therefore used to approximate the distribution of the original data, which can be expressed as:
\begin{equation}
\label{xdeq}
x_{deq} = \frac{x_{q}+p_{z}}{f_{s}}.
\end{equation}
Note that when using an activation function like Relu at the last layer of the encoder, the maximum and minimum \textcolor{black}{values} of feature $\mathbf{v}$ \textcolor{black}{may vary with} different inputs, leading to quantization results with non-identical distributions, making signal detection and interpretation at the decoder more challenging. \textcolor{black}{To enhance} the practicability of the MUD, the Tanh function is employed at the last layer of each semantic encoder to constrain the output in between $[-s+d,s+d]$, where $s$ is a multiplicative factor, $d$ is bias, and $0<d<s$. For each $x_{i} \in \mathbf{v}$, the constellation set of it after dequantization \textcolor{black}{may} be presented by $\mathcal{C}^{deq} = \{c_{1}, c_{2}, \ldots, c_{2^{m}}\}$. During the \textcolor{black}{process of} power normalization, the mean power of the modulated signal $\mathbf{s}$ \textcolor{black}{may} be calculated as follows:

\begin{align}\label{deqn_ex1a}
\overline{p}_{\mathbf{s}}=\sum_{c_{i}\in\mathcal{C}^{deq}}|\boldsymbol{\phi}^{mod}(c_{i})|^{2}p(c_{i}),
\end{align}
where $p(c_{i})$ is the proportion of point $c_{i}$ in $\mathbf{v}^{deq}$. The modulation process introduced in the previous subsection can be described as a linear transform $\boldsymbol{\phi}^{mod}(\mathbf{v^{deq}})=\mathbf{w}\mathbf{v^{deq}}+\mathbf{b}$, where $\mathbf{w}$ and $\mathbf{b}$ are weight and bias parameters contained in the dense layer. After normalization, each element in $\mathbf{s}$ is divided by $\overline{p}_{s}$. Constellations with a smaller amplitude may suffer more severely from noise and interference during \textcolor{black}{the} transmission. \textcolor{black}{By} contrast, those with larger amplitude are less distorted. Therefore, it is possible to realize unequal protection of the source information by controlling $p(c_{i}|x_{i}, \boldsymbol{\varepsilon}^{enc})$, which can be learned automatically during the training process.  For those features of minor importance, the encoder will learn to map them into constellations with a smaller value, whereas features of greater importance will have their magnitude increased in the forward propagation by the encoder. Since the distortion on the least important features has little impact on the outcome, they can simply be mapped into zero points. Additionally, zero points can \textcolor{black}{help to lower} the mean power $\overline{p}_{\mathbf{s}}$, \textcolor{black}{resulting} in \textcolor{black}{better} protection for critical features. \textcolor{black}{As demonstrated,} the zero point exists in the constellation point set $\mathcal{C}^{deq}$ as well \textcolor{black}{as with} the asymmetric quantization scheme. Please \textcolor{black}{refer to} Appendix \ref{appendix} for detail proofs. 
\subsection{Metrics, Loss Function, and Training Procedure}
\subsubsection{Metrics}
 \textcolor{black}{For} semantic communication, the performance metrics used to assess \textcolor{black}{the} transmission performance are \textcolor{black}{different from} those used in traditional bit-based communication such as bit-error-rate (BER) or block-error-rate (BLR). \textcolor{black}{And the} main concern of this paper is the transmission of image and text semantic information. For image transmission, structural similarity (SSIM)\cite{ssim} and peak signal-to-noise ratio (PSNR) are two widely adopted metrics for measuring the similarity of two images. SSIM primarily evaluates \textcolor{black}{the} two images in terms of luminance, contrast, and structure, \textcolor{black}{whilst the} PSNR \textcolor{black}{assesses} the relationship between the maximum value of the image matrix and the background noise.

\textcolor{black}{To measure} the sentence transmission performance, two metrics \textcolor{black}{from} \cite{deepsc} are adopted, which \textcolor{black}{include} bilingual evaluation understudy (BLEU) score and sentence similarity\cite{bert}. The BLEU score mainly measures the similarity at the word level, while \textcolor{black}{the} sentence similarity take account of both the context information of each word and the meaning of sentences.
\subsubsection{Loss Function}
The system can still be jointly optimized using a combined loss function even though each block's design is disjoint. As discussed in \cite{multiusercom}, separate design \textcolor{black}{will make} fine-tuning easier and faster \textcolor{black}{and enhance} the system's flexibility. The optimization goal for the image semantic codec is to minimize the distance between the original input image and the generated one, including MSE and SSIM between \textcolor{black}{the} two images. The detailed expression of loss function $\mathcal{L}^{I}$ for training LSCI is Eq. (14) in \cite{lsci}. 

For text transmission, the loss function $\mathcal{L}^{S}$ \textcolor{black}{refers to} the end-to-end reconstruction error between the original sentence $\mathbf{x}^{S}$ and the reconstructed sentence $\widetilde{\mathbf{x}}^{S}$. Since the input and output sentences of DeepSC can be \textcolor{black}{regarded} as a joint distribution of vocabulary, we \textcolor{black}{adopt} the loss function design \textcolor{black}{from} \cite{deepsc}, where cross-entropy (CE) is adopted for measuring the distance between \textcolor{black}{the} two sentences.

For the modem model, the loss function is the Euclidean distance between $\mathbf{v}^{deq}$ and $\widetilde{\mathbf{v}}^{deq}$:

\begin{align}\label{modloss}
\mathcal{L}_{F}^{mod}&=\mathbb{E}_{\mathbf{v}}\left[\left\|\mathbf{v}_{F}^{deq}-\widetilde{\mathbf{v}_{F}}^{deq}\right\|^{2}\right],\\
\mathcal{L}_{N}^{mod}&=\mathbb{E}_{\mathbf{v}}\left[\left\|\mathbf{v}_{F}^{deq}-\widetilde{\mathbf{v}_{F}}^{deq}\right\|^{2}+
\left\|\mathbf{v}_{N}^{deq}-\widetilde{\mathbf{v}_{N}}^{deq}\right\|^{2}\right]\notag,
\end{align}
which takes into account the impact of the channel as well as the interference from other users. 
\subsubsection{Training Procedure}
Gradient descent algorithm is chosen for training the whole system, where the training phase can be divided into two stages: (\uppercase\expandafter{\romannumeral1}) \textcolor{black}{training of} the modem model, and (\uppercase\expandafter{\romannumeral2}) \textcolor{black}{training of} the semantic codec. The first training stage is illustrated in Algorithm \ref{alg:alg1}. A dataset $\mathcal{D}^{1}=\left\{\left(\mathbf{v}_{N}^{deq}, \mathbf{v}_{F}^{deq}\right)\right\}$ for training is generated by sampling from the quantized constellation set $\mathcal{C}^{deq}$ with even probability $\frac{1}{2^{m}}$ for each element. In this way, the modulator model \textcolor{black}{could} generate a two-dimensional symbol constellation for every $c_{i}\in\mathcal{C}^{deq}$ while the demodulator model can learn to decode from the composite symbols under all pairing cases. After convergence, the trained models can be directly used for the second stage of training.

The second stage of training is illustrated in Algorithm \ref{alg:alg2}, where the image and text semantic codecs are trained simultaneously and \textcolor{black}{the} signal flow in one forward propagation follows the actual NOMA downlink transmission process. User-N is assumed to be an image-type user, and user-F is a textual data user. Note that though the two models can be trained separately, simultaneously training \textcolor{black}{has allowed} the model to learn from the estimation error, i.e., the difference between $\mathbf{v}^{deq}$ and $\widetilde{\mathbf{v}}^{deq}$ introduced by another user, \textcolor{black}{conferring} better \textcolor{black}{protection for} its own message.
\begin{breakablealgorithm}
\caption{Train Modem Model.}\label{alg:alg1}
\begin{algorithmic}
\STATE 
\STATE \textbf{Initialization:} \textcolor{black}{To randomly} initialize the model parameter \textcolor{black}{sets} $\boldsymbol{\theta}_{\boldsymbol{\phi}_{N}}^{0}$ and $\boldsymbol{\theta}_{\boldsymbol{\phi}_{F}}^{0}$. The training dataset $\mathcal{D}^{1}$, epoch number $E^{1}$ and batch size $B^{1}$

{\textsc{TRAIN}}$\left(\boldsymbol{\phi}_{N}^{mod},\boldsymbol{\phi}_{N}^{demod},\boldsymbol{\phi}_{F}^{mod},\boldsymbol{\phi}_{F}^{demod}\right)$
\FOR{$t=1$ to $E^{1}$}

\STATE \hspace{0.4cm} Select mini-batch data $\left\{\left(\mathbf{v}^{deq}_{N}, \mathbf{v}^{deq}_{F}\right)\right\}_{i=n}^{n+B^{1}}$ from $\mathcal{D}^{1}$ 

\STATE \hspace{0.4cm} Modulation:

\STATE \hspace{0.4cm} $\left\{\left(\mathbf{v}^{deq}_{N}, \mathbf{v}^{deq}_{F}\right)\right\}_{i=n}^{n+B^{1}}\stackrel{\boldsymbol{\phi}^{mod}}{\longrightarrow}\left\{\left(\mathbf{s}_{N}, \mathbf{s}_{F}\right)\right\}_{i=n}^{n+B^{1}}$ 

\STATE \hspace{0.4cm} Power normalization:

\STATE \hspace{0.5cm}$\left\{\left(\mathbf{s}_{N}, \mathbf{s}_{F}\right)\right\}_{i=n}^{n+B^{1}}\longrightarrow \left\{\left(\mathbf{s}_{N}^{norm}, \mathbf{s}_{F}^{norm}\right)\right\}_{i=n}^{n+B^{1}}$

\STATE \hspace{0.4cm} Superimposed coding:

\STATE \hspace{0.4cm}$\{(\mathbf{s}_{N}^{norm}, \mathbf{s}_{F}^{norm})\}_{i=n}^{n+B^{1}}\longrightarrow \{\mathbf{x}_{sc}\}_{i=n}^{n+B^{1}}$

\STATE \hspace{0.4cm} Transmitted through channel:

\STATE \hspace{0.5cm}$\{\mathbf{x}_{sc}\}_{i=n}^{n+B^{1}}\longrightarrow \{(\mathbf{y}_{N}, \mathbf{y}_{F})\}_{i=n}^{n+B^{1}}$

\STATE \hspace{0.4cm} Demodulation:

\STATE \hspace{0.4cm}  $\{(\mathbf{y}_{N}, \mathbf{y}_{F})\}_{i=n}^{n+B^{1}}\stackrel{\boldsymbol{\phi}^{demod}}{\longrightarrow} \left\{\left(\widetilde{\mathbf{v}}^{deq}_{N}, \widetilde{\mathbf{v}}^{deq}_{F}\right)\right\}_{i=n}^{n+B^{1}}$

\STATE \hspace{0.4cm}  Compute loss $\mathcal{L}_{N}^{mod}$ and $\mathcal{L}_{F}^{mod}$ based on (\ref{modloss}) with $\left\{\left(\mathbf{v}^{deq}_{N}, \mathbf{v}^{deq}_{F}\right)\right\}_{i=n}^{n+B^{1}}$ and $\left\{\left(\widetilde{\mathbf{v}}^{deq}_{N}, \widetilde{\mathbf{v}}^{deq}_{F}\right)\right\}_{i=n}^{n+B^{1}}$

\STATE \hspace{0.4cm} Update $\boldsymbol{\theta}_{\boldsymbol{\phi}_{N}}^{t}$ and $\boldsymbol{\theta}_{\boldsymbol{\phi}_{F}}^{t}$ through gradient descent with $\mathcal{L}_{N}^{mod}$ and $\mathcal{L}_{F}^{mod}$
\ENDFOR
\STATE \textbf{Return} Trained parameters $\boldsymbol{\theta}_{\boldsymbol{\phi}_{N}}^{E^{1}}$ and $\boldsymbol{\theta}_{\boldsymbol{\phi}_{F}}^{E^{1}}$
\end{algorithmic}
\label{alg1}
\end{breakablealgorithm}

\begin{breakablealgorithm}
\caption{Train Semantic Codec.}\label{alg:alg2}
\begin{algorithmic}
\STATE 
\STATE \textbf{Initialization:} \textcolor{black}{To randomly} initialize the model parameter \textcolor{black}{sets} $\left\{\boldsymbol{\theta}_{\boldsymbol{\varepsilon}_{N}^{I}}\right\}^{0}$ and $\left\{\boldsymbol{\theta}_{\boldsymbol{\varepsilon}_{F}^{S}}\right\}^{0}$. The training dataset $\mathcal{D}^{2}$, epoch number $E^{2}$ and batch size $B^{2}$

{\textsc{TRAIN}}$\left(\boldsymbol{\varepsilon}^{I}, \boldsymbol{\varepsilon}^{S}\right)$
\FOR{$t=1$ to $E^{2}$}

\STATE \hspace{0.5cm} Select mini-batch data $\left\{\left(\mathbf{x}^{I}_{N}, \mathbf{x}^{S}_{F}\right)\right\}_{i=n}^{n+B^{2}}$ from $\mathcal{D}^{2}$ 

\STATE \hspace{0.5cm} Semantic encoding:

\STATE \hspace{0.5cm} $\left\{\left(\mathbf{x}^{I}_{N}, \mathbf{x}^{S}_{F}\right)\right\}_{i=n}^{n+B^{2}}\stackrel{\boldsymbol{\varepsilon}^{enc}}{\longrightarrow}\left\{\left(\mathbf{v}^{I}_{N}, \mathbf{v}^{S}_{F}\right)\right\}_{i=n}^{n+B^{2}}$ 

\STATE \hspace{0.5cm} Asymmetric quantization:

\STATE \hspace{0.5cm} $\left\{\left(\mathbf{v}^{I}_{N}, \mathbf{v}^{S}_{F}\right)\right\}_{i=n}^{n+B^{2}}\stackrel{\mathcal{Q}}{\longrightarrow}\left\{\left(\mathbf{v}^{deq,I}_{N}, \mathbf{v}^{deq,S}_{F}\right)\right\}_{i=n}^{n+B^{2}}$ 

\STATE \hspace{0.5cm} Modulation:

\STATE \hspace{0.5cm} $\left\{\left(\mathbf{v}^{deq,I}_{N}, \mathbf{v}^{deq,S}_{F}\right)\right\}_{i=n}^{n+B^{2}}\stackrel{\boldsymbol{\phi}^{mod}}{\longrightarrow}\left\{\left(\mathbf{s}^{I}_{N}, \mathbf{s}^{S}_{F}\right)\right\}_{i=n}^{n+B^{2}}$ 

\STATE \hspace{0.5cm} Power normalization:

\STATE \hspace{0.5cm}$\left\{\left(\mathbf{s}^{I}_{N}, \mathbf{s}^{S}_{F}\right)\right\}_{i=n}^{n+B^{2}}\longrightarrow \left\{\left(\mathbf{s}^{norm,I}_{N}, \mathbf{s}^{norm,S}_{F}\right)\right\}_{i=n}^{n+B^{2}}$

\STATE \hspace{0.5cm} Superimposed coding:

\STATE \hspace{0.5cm}$\left\{\left(\mathbf{s}^{norm,I}_{N}, \mathbf{s}^{norm,S}_{F}\right)\right\}_{i=n}^{n+B^{2}}\longrightarrow \{\mathbf{x}_{sc}\}_{i=n}^{n+B^{2}}$

\STATE \hspace{0.5cm} Transmitted through channel:

\STATE \hspace{0.5cm}$\{\mathbf{x}_{sc}\}_{i=n}^{n+B^{2}}\longrightarrow \{(\mathbf{y}_{N}, \mathbf{y}_{F})\}_{i=n}^{n+B^{2}}$

\STATE \hspace{0.5cm} Demodulation:

\STATE \hspace{0.5cm}  $\{(\mathbf{y}_{N}, \mathbf{y}_{F})\}_{i=n}^{n+B^{2}}\stackrel{\boldsymbol{\phi}^{demod}}{\longrightarrow} \left\{\left(\widetilde{\mathbf{v}}^{deq,I}_{N}, \widetilde{\mathbf{v}}^{deq,S}_{F}\right)\right\}_{i=n}^{n+B^{2}}$

\STATE \hspace{0.5cm} Semantic Decoding:

\STATE \hspace{0.5cm}  $\left\{\left(\widetilde{\mathbf{v}}^{deq,I}_{N}, \widetilde{\mathbf{v}}^{deq,S}_{F}\right)\right\}_{i=n}^{n+B^{2}}\stackrel{\boldsymbol{\varepsilon}^{dec}}{\longrightarrow} \left\{\left(\widetilde{\mathbf{x}}^{I}_{N}, \widetilde{\mathbf{x}}^{S}_{F}\right)\right\}_{i=n}^{n+B^{2}}$

\STATE \hspace{0.5cm}  Compute loss $\mathcal{L}_{N}^{I}$ and $\mathcal{L}_{F}^{S}$ with $\left\{\left(\mathbf{x}^{I}_{N}, \mathbf{x}^{S}_{F}\right)\right\}_{i=n}^{n+B^{2}}$ and $(\left\{\left(\widetilde{\mathbf{x}}^{I}_{N}, \widetilde{\mathbf{x}}^{S}_{F}\right)\right\}_{i=n}^{n+B^{2}}$

\STATE \hspace{0.5cm} Update $\left\{\boldsymbol{\theta}_{\boldsymbol{\varepsilon}^{I}}\right\}^{t}$ and $\left\{\boldsymbol{\theta}_{\boldsymbol{\varepsilon}^{S}}\right\}^{t}$ through gradient descent with $\mathcal{L}_{N}^{I}$ and $\mathcal{L}_{F}^{S}$
\ENDFOR
\STATE \textbf{Return} Trained parameters $\left\{\boldsymbol{\theta}_{\boldsymbol{\varepsilon}}^{I}\right\}^{E^{2}}$ and $\left\{\boldsymbol{\theta}_{\boldsymbol{\varepsilon}}^{S}\right\}^{E^{2}}$
\end{algorithmic}
\label{alg1}
\end{breakablealgorithm} 
\section{Performance Analysis}
Similar \textcolor{black}{with} the analysis \textcolor{black}{of} Semi-NOMA in \cite{seminoma}, in this section, the theoretical transmission performance of \textcolor{black}{the} NOMASC is analyzed and compared \textcolor{black}{with} the \textcolor{black}{scenario of} OMA based semantic communication. The definitions \textcolor{black}{for} text and image semantic transmission rates are given first, and the rate region and power region of \textcolor{black}{the} NOMASC are characterized \textcolor{black}{by using} the defined semantic transmission rate and exhaustive search method.
\subsection{Semantic Transmission Rate}
It is important to define the semantic transmission rate (S-Rate) for network-level performance analysis and optimization. The S-Rate for text semantic communication \textcolor{black}{is adopted from} \cite{semanticrate}. For a text dataset containing sentences with average length $L_{S}$ and average semantic information per sentence $I_{S}$ (measured in \textcolor{black}{semantic unit per second (suts)}), the corresponding text S-Rate can be represented as follows:
\begin{align}\label{deqn_ex1a}
\Gamma_{S} = \frac{WI_{S}}{KL_{S}}\boldsymbol{\xi}_{K}^{S}(\gamma)\ (suts/s),
\end{align}
where $W$ denotes the bandwidth of \textcolor{black}{the} transmission channel, \textcolor{black}{and it also equals to the symbol rate for passband transmission}. The text S-Rate mainly \textcolor{black}{comprises} two parts, the fraction term refers to the average semantic information \textcolor{black}{which} each semantic symbol conveys, and the second term \textcolor{black}{refers to} the sentence similarity, which is mainly based on the number of symbols per word $K$ and transmission SNR $\gamma$. This definition couples the semantic rate with the transmission accuracy and represents the average amount of successfully transmitted semantic information per second. Similarly, the S-Rate of image semantic transmission \textcolor{black}{is} defined as follows:
\begin{align}\label{deqn_ex1a}
\Gamma_{I} = \frac{WI_{I}}{CrL_{I}}\boldsymbol{\xi}_{Cr}^{I}(\gamma) \ (suts/s),
\end{align}
where $I_{I}$ is the average amount of semantic information carried in a single image and $L_{I}$ is the average pixel number in a single image (e.g., for an image with height $H$, width $W$ and channel number $C$, $L_{I}$ is $H*W*C$). For a fixed compression ratio $Cr$, $CrL_{I}$ represents the average number of semantic symbols per image. The semantic metric employed in this definition is SSIM.
\subsection{Rate Region}
For the two-user NOMA downlink scenario considered in this paper, the received SNR of user-N and user-F can be denoted as:
\begin{equation}
\begin{aligned}\label{24}
\gamma_{N} &= \frac{\rho_{N}P_{max}|\mathbf{h}_{N}|^{2}}{\sigma_{N}^{2}W}, \\ 
\gamma_{F} &= \frac{\rho_{F}P_{max}|\mathbf{h}_{F}|^{2}}{\rho_{N}P_{max}|\mathbf{h}_{F}|^{2}+\sigma_{F}^{2}W}.
\end{aligned}
\end{equation}
User-F decodes its message \textcolor{black}{by} treating the interference from user-N as noise. And user-N is able to detect the signal of both users. Therefore, the signal of user-F is not regarded as an interference for user-N. For the characterization of the rate region and power region, it is assumed that user-N requires textual data, while user-F requires image-type data for simplicity. The same analysis can be done for any other pairing situations. The downlink S-Rate of each user can be \textcolor{black}{calculated} as:
\begin{equation}
\begin{aligned}\label{deqn_ex1a}
\Gamma_{N} &= \frac{WI_{S}}{KL_{S}}\boldsymbol{\xi}_{K}^{S}\left(\frac{\rho_{N}P_{max}|\mathbf{h}_{N}|^{2}}{\sigma_{N}^{2}W}\right),\\
\Gamma_{F} &= \frac{WI_{I}}{CrL_{I}}\boldsymbol{\xi}_{Cr}^{I}\left(\frac{\rho_{F}P_{max}|\mathbf{h}_{F}|^{2}}{\rho_{N}P_{max}|\mathbf{h}_{F}|^{2}+\sigma_{F}^{2}W}\right).
\end{aligned}
\end{equation}
Note that the characterization of the NOMA rate region is carried out under the \textcolor{black}{requirements for} the semantic accuracy $\boldsymbol{\xi}_{K}^{S, req}$ and $\boldsymbol{\xi}_{Cr}^{I, req}$, since the concern of semantic users may be not only the transmission rate but also the accuracy of contents. And in order to satisfy the accuracy requirement, the minimum power allocation factor can be calculated as follows:
\begin{equation}
\begin{aligned}
\label{deqn_ex1a}
\rho_{N}^{min} &= \left(\boldsymbol{\xi}_{K}^{S}\right)^{-1}\left(\boldsymbol{\xi}_{K}^{S,req}\right)\frac{W\sigma^{2}_{N}}{P_{max}|h_{N}|^{2}},\\
\rho_{F}^{min} &= \frac{\left(\boldsymbol{\xi}_{Cr}^{I}\right)^{-1}\left(\boldsymbol{\xi}_{Cr}^{I,req}\right)\left(|\mathbf{h}_{F}|^{2}+W\sigma^{2}_{F}/P_{max}\right)}{|\mathbf{h}_{F}|^{2}\left[\left(\boldsymbol{\xi}_{Cr}^{I}\right)^{-1}\left(\boldsymbol{\xi}_{Cr}^{I,req}\right)+1\right]},
\end{aligned}
\end{equation}
where each point in the rate region should satisfy $\rho_{N}+\rho_{F}=1$ to achieve the maximum possible $\Gamma_{N}$ and $\Gamma_{F}$. Besides, $\rho_{N}^{min}+\rho_{F}^{min}<1$ should be met, otherwise the rate region is $\emptyset$. Each point in the rate region can be calculated by fixing $\Gamma_{N}$ and computing the maximum $\Gamma_{F}$. For each $\Gamma_{N}^{min} = \frac{WI_{S}}{KL_{S}}\boldsymbol{\xi}_{K}^{S,req}\leq \overline{\Gamma_{N}} \leq \Gamma_{F}^{max} = \frac{WI_{S}}{KL_{S}}\boldsymbol{\xi}_{K}^{S}\left(\frac{P_{max}|\mathbf{h}_{N}|^{2}}{\sigma_{N}^{2}W}\right)$, the minimum power factor for user-N is:
\begin{align}\label{deqn_ex1a}
\overline{\rho_{N}}^{min}=\left(\boldsymbol{\xi}_{K}^{S}\right)^{-1}\left(\frac{\overline{\Gamma_{N}}KL_{S}}{WI_{S}}\right)\frac{W\sigma_{N}^{2}}{P_{max}|h_{N}|^{2}}.
\end{align}
By allocating the remaining power $\left(1-\overline{\rho_{N}}^{min}\right)P_{max}$ to user-F, the desired maximum $\Gamma_{F}$ can be obtained. 

Under the orthogonal multiple access (OMA) scenarios, each user is given two degrees of freedom: power and bandwidth. The transmission path of each user is orthogonal to each other, thus there will be no inter-user interference. The received SNRs of user-N and user-F are \textcolor{black}{calculated} as follows:
\begin{align}\label{deqn_ex1a}
\gamma_{i} = \frac{\rho_{i}P_{max}|\mathbf{h}_{i}|^{2}}{\sigma_{i}^{2}W_{i}}, \ i\in\{N,F\}.
\end{align}
And the downlink S-Rate for \textcolor{black}{the} two users is:
\begin{equation} 
\begin{aligned}\label{deqn_ex1a}
\Gamma_{N} = \frac{W_{N}I_{S}}{KL_{S}}\boldsymbol{\xi}_{K}^{S}\left(\frac{\rho_{N}P_{max}|\mathbf{h}_{N}|^{2}}{\sigma_{N}^{2}W_{N}}\right),\\
\Gamma_{F} = \frac{W_{F}I_{I}}{CrL_{I}}\boldsymbol{\xi}_{Cr}^{I}\left(\frac{\rho_{F}P_{max}|\mathbf{h}_{F}|^{2}}{\sigma_{F}^{2}W_{F}}\right),
\end{aligned}
\end{equation}
where $W_{N}$ and $W_{F}$ are the channel bandwidth allocated for user-N and user-F, respectively. Two corner points in the rate region of OMA are $\left(\frac{WI_{S}}{KL_{S}}\boldsymbol{\xi}_{K}^{S}\left(\frac{P_{max}|\mathbf{h}_{N}|^{2}}{\sigma_{N}^{2}W_{N}}\right), 0\right)$ and $\left(0, \frac{WI_{I}}{CrL_{I}}\boldsymbol{\xi}_{Cr}^{I}\left(\frac{P_{max}|\mathbf{h}_{F}|^{2}}{\sigma_{F}^{2}W_{F}}\right)\right)$, which can be achieved by \textcolor{black}{allowing} one user complete access to the transmit power and bandwidth. The remaining points in the rate region can be characterized by solving the following problem:
%
\begin{small}
\begin{align}\label{deqn_ex1a}
\max\limits_{W_{N},\rho_{N}}\frac{\left(W-W_{N}\right)I_{I}}{CrL_{I}}&\boldsymbol{\xi}_{Cr}^{I}\left(\frac{(1-\rho_{N})P_{max}|\mathbf{h}_{F}|^{2}}{\sigma_{F}^{2}(W-W_{N})}\right) \\
s.t. \quad\frac{W_{N}I_{S}}{KL_{S}}&\boldsymbol{\xi}_{K}^{S}\left(\frac{\rho_{N}P_{max}|\mathbf{h}_{N}|^{2}}{\sigma_{N}^{2}W_{N}}\right)\geq \overline{\Gamma_{N}},\tag{21a}\label{opt1a}\\
&\boldsymbol{\xi}_{K}^{S}\left(\frac{\rho_{N}P_{max}|\mathbf{h}_{N}|^{2}}{\sigma_{N}^{2}W_{N}}\right)\geq\boldsymbol{\xi}_{K}^{S,req},\tag{21b}\label{opt1b}\\
&\boldsymbol{\xi}_{Cr}^{I}\left(\frac{(1-\rho_{N})P_{max}|\mathbf{h}_{F}|^{2}}{\sigma_{F}^{2}(W-W_{N})}\right)\geq\boldsymbol{\xi}_{Cr}^{I, req},\tag{21c}\label{opt1c}\\
&0 \leq W_{N} \leq W,\tag{21d}\\
&0 \leq\rho_{N}\leq 1. \tag{21e}
\end{align}
\end{small}
where all points in \textcolor{black}{the} OMA rate region satisfy $\rho_{N}+\rho_{F}=1$ and $W_{N} + W_{F} = W$, therefore $\rho_{F}$ and $W_{F}$ can be directly replaced by $\left(1-\rho_{N}\right)$ and $\left(W-W_{N}\right)$ for simplicity, $\overline{\Gamma_{N}}$ is the rate requirement for user-N. Here we \textcolor{black}{have made} the assumption that \textcolor{black}{the two accuracy constraints} $\boldsymbol{\xi}_{K}^{S,req}$ and $\boldsymbol{\xi}_{Cr}^{I, req}$ can still be satisfied in the two corner points using $\lim\limits_{W\to0,\ \rho\to0} \boldsymbol{\xi}(\frac{\rho P_{max}|\mathbf{h}|^{2}}{\sigma^{2}W})\geq0 $. Constraint (\ref{opt1a}) can be transformed into $\boldsymbol{\xi}_{K}^{S}\left(\frac{\rho_{N}P_{max}|\mathbf{h}_{N}|^{2}}{\sigma_{N}^{2}W_{N}}\right) \geq \frac{\overline{\Gamma_{N}}KL_{S}}{W_{N}I_{S}}$. Since $\boldsymbol{\xi}_{K}^{S} \leq 1$, the lower bound of $W_{N}$ can be \textcolor{black}{determined} as: $W_{N} \geq \frac{\overline{\Gamma_{N}}KL_{S}}{I_{S}}$. And for $\forall\ {\overline{W_{N}}}\in\left[\frac{\overline{\Gamma_{N}}KL_{S}}{I_{S}}, W\right)$, the lower bound of $\rho_{N}$ can be \textcolor{black}{determined} using (\ref{opt1a}) and (\ref{opt1b}):
\begin{align}\label{deqn_ex1a}
\rho_{N}^{low} = \max\Bigg\{0, \frac{\sigma_{N}^{2}W_{N}}{|\mathbf{h}_{N}|^{2}P_{max}}\left(\boldsymbol{\xi}_{K}^{S}\right)^{-1}\left(\frac{\overline{\Gamma_{N}}KL_{S}}{\overline{W_{N}}I_{S}}\right), \notag\\
\frac{\sigma_{N}^{2}W_{N}}{|\mathbf{h}_{N}|^{2}P_{max}}\left(\boldsymbol{\xi}_{K}^{S}\right)^{-1}\left(\boldsymbol{\xi}_{K}^{S,req}\right)\Bigg\}.
\end{align}
And the maximum of $\rho_{N}$ can be computed using (\ref{opt1c}):
\begin{align}\label{deqn_ex1a}
\rho_{N}^{up} = \min\left\{\frac{\sigma_{N}^{2}(W-W_{N})\left(\boldsymbol{\xi}_{K}^{S}\right)^{-1}\left(\boldsymbol{\xi}_{K}^{S,req}\right)}{|\mathbf{h}_{F}|^{2}P_{max}}+1, 1\right\}.
\end{align}
Note that $\rho_{N}^{low}\leq1$, $\rho_{N}^{up}\geq0$ and $\rho_{N}^{low}\leq\rho_{N}^{up}$ should be satisfied. Therefore, for $\forall \ \overline{\Gamma_{N}}\in\left[0, \ \frac{WI_{S}}{KL_{S}}\boldsymbol{\xi}_{K}^{S}\left(\frac{P_{max}|\mathbf{h}_{N}|^{2}}{\sigma_{N}^{2}W}\right)\right]$, the corresponding maximum $\Gamma_{F}$ can be obtained by carrying out \textcolor{black}{an} exhaustive search over $W_{N}\in\left[\frac{\overline{\Gamma_{N}}KL_{S}}{I_{S}}, W\right)$:
\begin{small}
\begin{align}\label{36}
\Gamma_{F}^{max}=\mathop{argmax}\limits_{\begin{tiny}W_{N}\end{tiny}}\frac{(W-W_{N})I_{I}}{CrL_{I}}\boldsymbol{\xi}_{Cr}^{I}\left(\frac{\left(1-\rho_{N}^{low}\right)P_{max}|\mathbf{h}_{F}|^{2}}{\sigma_{F}^{2}\left(W-W_{N}\right)}\right).
\end{align}
\end{small}

\subsection{Power Region}
The power region can be characterized by searching for the minimum power required for satisfying the S-Rate constraint $\Gamma_{N}^{req}, \ \Gamma_{F}^{req}$ and accuracy constraint $\boldsymbol{\xi}_{K}^{S, req},\ \boldsymbol{\xi}_{Cr}^{I,req}$. Each point in the power region can be obtained by solving the following optimization problem:
\begin{align}\label{37}
\min\limits_{0<W_{N}<W,\rho_{N}\geq0, \rho_{F}\geq0}\quad \left(\rho_{N}+\rho_{F}\right)P_{max}\\
s.t.\quad \frac{W_{N}I_{S}}{KL_{S}}\boldsymbol{\xi}_{K}^{S}\left(\frac{\rho_{N}P_{max}|\mathbf{h}_{N}|^{2}}{W_{N}\sigma_{N}^{2}}\right)\geq \Gamma_{N}^{req},\tag{25a}\label{opt2a}\\
\frac{(W-W_{N})I_{I}}{CrL_{I}}\boldsymbol{\xi}_{Cr}^{I}\left(\frac{\rho_{F}P_{max}|\mathbf{h}_{F}|^{2}}{(W-W_{N})\sigma_{F}^{2}}\right)\geq \Gamma_{F}^{req},\tag{25b}\label{opt2b}\\
\boldsymbol{\xi}_{K}^{S}\left(\frac{\rho_{N}P_{max}|\mathbf{h}_{N}|^{2}}{W_{N}\sigma_{N}^{2}}\right)\geq \boldsymbol{\xi}_{K}^{S, req},\tag{25c}\label{opt2c}\\
\boldsymbol{\xi}_{Cr}^{I}\left(\frac{\rho_{F}P_{max}|\mathbf{h}_{F}|^{2}}{(W-W_{N})\sigma_{F}^{2}}\right)\geq \boldsymbol{\xi}_{Cr}^{I,req},\tag{25d}\label{opt2d}
\end{align}
where (\ref{opt2a}) can be transformed into $\boldsymbol{\xi}_{K}^{S}\left(\frac{\rho_{N}P_{max}|\mathbf{h}_{N}|^{2}}{W_{N}\sigma_{N}^{2}}\right)\geq\frac{\Gamma_{N}^{req}KL_{S}}{W_{N}I_{S}}$. Since $\frac{\Gamma_{N}^{req}KL_{S}}{W_{N}I_{S}}\leq1$, the range of $W_{N}$ can be computed as $W_{N}\geq \frac{\Gamma_{N}^{req}KL_{S}}{I_{S}}$. The minimum of $\rho_{N}$ can be obtained using (\ref{opt2a}) and (\ref{opt2c}):
\begin{align}\label{deqn_ex1a}
\rho_{N}^{min} = \max\bigg\{0, \frac{\sigma_{N}^{2}W_{N}}{P_{max}|\mathbf{h}_{N}|^{2}}\left(\boldsymbol{\xi}_{K}^{S}\right)^{-1}\left(\frac{\Gamma_{N}^{req}KL_{S}}{W_{N}I_{S}}\right),\notag \\
\frac{\sigma_{N}^{2}W_{N}}{P_{max}|\mathbf{h}_{N}|^{2}}\left(\boldsymbol{\xi}_{K}^{S}\right)^{-1}\left(\boldsymbol{\xi}_{K}^{S,req}\right)\bigg\},
\end{align}
while the minimum of $\rho_{F}$ can be obtained using (\ref{opt2b}) and (\ref{opt2d}):
\begin{align}\label{deqn_ex1a}
\rho_{F}^{min} = \max\bigg\{0, \frac{\sigma_{F}^{2}(W-W_{N})}{P_{max}|\mathbf{h}_{F}|^{2}}\left(\boldsymbol{\xi}_{Cr}^{I}\right)^{-1}\left(\frac{\Gamma_{F}^{req}CrL_{I}}{(W-W_{N})I_{I}}\right),\notag\\
\frac{\sigma_{F}^{2}(W-W_{N})}{P_{max}|\mathbf{h}_{F}|^{2}}\left(\boldsymbol{\xi}_{Cr}^{I}\right)^{-1}\left(\boldsymbol{\xi}_{Cr}^{I,req}\right)\bigg\},
\end{align}
where $\rho_{N}^{min}\leq1, \rho_{F}^{min}\leq1, \rho_{N}^{min}+\rho_{F}^{min}\leq1$ also need to be satisfied. The minimum power $P_{total}^{min}$ under the S-Rate requirement and accuracy requirement can be obtained through an exhaustive search over the range of $W_{N}\in\left[\frac{\Gamma_{N}^{req}KL_{S}}{I_{S}},W\right)$:
\begin{align}\label{deqn_ex1a}
P_{total}^{min}=\mathop{argmin}\limits_{W_{N}}\left[\rho_{N}^{min}(W_{N})+\rho_{F}^{min}(W_{N})\right]P_{max}.
\end{align}
As for the NOMA scenario, the power region can be characterized by solving the following optimization problem:
\begin{align}\label{deqn_ex1a}
&\min\limits_{\rho_{N}\geq0, \rho_{F}\geq0}\quad (\rho_{N}+\rho_{F})P_{max}\\
s.t.\quad \frac{WI_{S}}{KL_{S}}&\boldsymbol{\xi}_{K}^{S}\left(\frac{\rho_{N}P_{max}|\mathbf{h}_{N}|^{2}}{W\sigma_{N}^{2}}\right)\geq \Gamma_{N}^{req},\tag{29}\label{opt3a}\\
\frac{WI_{I}}{CrL_{I}}&\boldsymbol{\xi}_{Cr}^{I}\left(\frac{\rho_{F}P_{max}|\mathbf{h}_{F}|^{2}}{W\sigma_{F}^{2}+\rho_{N}P_{max}|\mathbf{h}_{F}|^{2}}\right)\geq \Gamma_{F}^{req},\tag{29b}\label{opt3b}\\
&\boldsymbol{\xi}_{K}^{S}\left(\frac{\rho_{N}P_{max}|\mathbf{h}_{N}|^{2}}{W\sigma_{N}^{2}}\right)\geq \boldsymbol{\xi}_{K}^{S,req},\tag{29c}\label{opt3c}\\
&\boldsymbol{\xi}_{Cr}^{I}\left(\frac{\rho_{F}P_{max}|\mathbf{h}_{F}|^{2}}{W\sigma_{F}^{2}+\rho_{N}P_{max}|\mathbf{h}_{F}|^{2}}\right)\geq \boldsymbol{\xi}_{Cr}^{I,req}.\tag{29d}\label{opt3d}
\end{align}
There is only one degree of freedom for each user under the NOMA scenario and points in the NOMA power region can also be computed by exhaustive search. The minimum of $\rho_{N}$ can be calculated using (\ref{opt3a}) and (\ref{opt3c}) as follows:
\begin{align}\label{deqn_ex1a}
\rho_{N}^{min} = max\bigg\{\frac{W\sigma_{N}^{2}}{P_{max}|h_{N}|^{2}}\left(\boldsymbol{\xi}_{K}^{S}\right)^{-1}\left(\frac{\Gamma_{N}^{req}KL_{S}}{WI_{S}}\right), \notag \\
\frac{W\sigma_{N}^{2}}{P_{max}|h_{N}|^{2}}\left(\boldsymbol{\xi}_{K}^{S}\right)^{-1}\left(\boldsymbol{\xi}_{K}^{S,req}\right), 0\bigg\}.
\end{align}
And for a fixed $\overline{\rho_{N}} \in \left[\rho_{N}^{min}, 1 \right]$, the minimum of $\rho_{F}$ can be calculated using (\ref{opt3b}) and (\ref{opt3d}) as follows:
\begin{small}
\begin{align}\label{deqn_ex1a}
\rho_{F}^{min} = max\bigg\{\left(\boldsymbol{\xi}_{Cr}^{I}\right)^{-1}\left(\boldsymbol{\xi}_{Cr}^{I,req}\right)\left(\frac{W\sigma_{F}^{2}}{P_{max}|\mathbf{h}_{F}|^{2}}+\overline{\rho_{N}}\right), \notag\\ \left(\boldsymbol{\xi}_{Cr}^{I}\right)^{-1}\left(\frac{\Gamma_{F}^{req}CrL_{I}}{WI_{I}}\right)\left(\frac{W\sigma_{F}^{2}}{P_{max}|\mathbf{h}_{F}|^{2}}+\overline{\rho_{N}}\right), 0\bigg\},
\end{align}
\end{small}
where $\rho_{N}^{min}<1,\ \rho_{F}^{min}<1$ and $\rho_{F}^{min}+\rho_{F}^{min}\leq1$ also need to be satisfied. The minimum transmission power $P_{total}^{min}$ can be obtained \textcolor{black}{through} an exhaustive search over $\rho_{N}$:

\begin{align}\label{44}
P_{total}^{min}=\mathop{argmin}\limits_{\overline{\rho_{N}}\in\left[\rho_{N}^{min}, 1 \right]}\left[\overline{\rho_{N}}+\rho_{F}^{min}\left(\overline{\rho_{N}}\right)\right]P_{max}.
\end{align}

\section{Experiments and Numerical Results}
Simulations are carried out in this part to \textcolor{black}{assess} the effectiveness of the proposed NOMASC system. The quantization and modulation processes are illustrated first. Meanwhile, a series of experiments are conducted under different channel types and settings, \textcolor{black}{to compare the} NOMASC \textcolor{black}{with} several baseline methods to demonstrate its superiority and robustness.
\subsection{Datasets and Experiment Settings}
Three datasets are used for \textcolor{black}{the} evaluation, including MNIST\cite{mnist}(gray-scale handwritten character digits $28 \times 28$ images), CIFAR10\cite{cifar}($32 \times 32$ images with different classes) and the proceedings of the European Parliament (Europarl)\cite{europarl}(consists of about 2.0 million sentences).
And to verify the capability of \textcolor{black}{the} NOMASC for serving user pairs with \textcolor{black}{various} modalities and datasets, \textcolor{black}{its} transmission performance is evaluated under three cases including:
\begin{itemize}
    \item [1)]
    Cifar\&Cifar: user-N and user-F are \textcolor{black}{both} Cifar users; this case corresponds to the scenario that \textcolor{black}{the} users with identical modalities and datasets are paired and served.
    \item [2)]
    Cifar\&MNIST: user-N is a Cifar user, and user-F is a MNIST user; this case corresponds to the scenario where users with identical modalities but distinct datasets are paired and served.
    \item [3)]
    Cifar\&Europarl: user-N is a Cifar user, and user-F is a Europarl user; this case corresponds to the scenario that users with non-identical modalities and datasets are paired and served. 
\end{itemize}
Despite both the CIFAR user and the MNIST user require image-type semantic information, the structure of their models is not entirely the same since the image's resolution is different. For MNIST users, the semantic encoder and decoder each \textcolor{black}{contains} 3 convolution layers and 6 residual blocks. For Cifar users, \textcolor{black}{each} encoder and decoder model has 7 convolution layers and 6 residual blocks. Each part of the discriminator for MNIST and Cifar users contains two convolution layers and a batch-normalization layer. The semantic encoder and decoder in DeepSC both have 4 transformer blocks with 8 attention heads, and the width of each transformer layer is 512. For each word in the sentence, it is embedded into a vector with a length of 128. The optimizer used for training the semantic codec is Adam with a learning rate of $1e-4$, whereas the SGD optimizer with a learning rate of $0.1$ is used to train the modem model. Some other parameters used in the simulation are \textcolor{black}{listed} in Table \ref{tab:table1}.
\begin{table}[!t]
\caption{Parameter setting in \textcolor{black}{the} training and testing \textcolor{black}{processes}\label{tab:table1}}
\centering
\begin{tabular}{|c|c||c|c| }
\hline
$\rho_{N}$ & 0.3 & $\rho_{F}$ & 0.7\\
\hline
$m_{N}$ & 2 & $m_{F}$ & 2\\
\hline
$SNR_{N}^{train}$ & 14\ dB &$ SNR_{F}^{train}$ & 6\ dB\\
\hline
$SNR_{N}^{test}$ & [0,28]\ dB & $SNR_{F}^{test}$ & [-8,20]\ dB\\
\hline
$E^{1}$ & 2000 & $E^{2}$ & 200 \\
\hline
$B^{1}$ & 4 &$ B^{2}$ & 16\\
\hline
$P_{max}$ & 1\ MW & $W$ & 1\ MHz\\
\hline
$s$ & 5 & $d$ & 1\\
\hline
$Cr^{Cifar}$ & 0.33 & $Cr^{MNIST}$ & 0.33\\
\hline
$K$ & 128 &  & \\
\hline
\end{tabular}
\end{table}
\subsection{Comparison Schemes}
To validate the advantages of \textcolor{black}{the} NOMASC, several other baseline methods \textcolor{black}{have} also \textcolor{black}{been} implemented for \textcolor{black}{the} comparison:
\begin{itemize}
  \item [1)]
    Conventional Scheme: \textcolor{black}{It performs the} source coding and channel coding separately. JPEG and LDPC are employed for \textcolor{black}{the} source and channel coding of \textcolor{black}{the} image-type sources, while for \textcolor{black}{the} textual sources, Huffman and Turbo \textcolor{black}{codes} are used. The LDPC code employs the DVB-S.2 standard, $\frac{1}{2}$ code rate. The code rate for Turbo code is set to $\frac{1}{3}$. The modulation scheme is QAM for converting bits into symbols. For multi-user detection, the SIC algorithm is employed.
  \item [2)]
    Hybrid Scheme: \textcolor{black}{It replaces} the modem model part of the NOMASC with QAM and SIC, where the output of the asymmetric quantizer is followed by a binary mapping to the bit stream based on the index of each element in the quantization set. \textcolor{black}{Subsequently}, QAM is carried out to map the bit stream to symbols, and the rest of the process is identical to the conventional scheme. 
  \item [3)]
    DeepJSCC-Q\cite{jsccq}: An end-to-end JSCC scheme with confined channel input and limited constellation points can be used for \textcolor{black}{the transmission}. Soft-to-hard quantization is used in \textcolor{black}{the} DeepJSCC-Q for mapping each element of \textcolor{black}{the} encoder output to its closest constellation point. The semantic codec and modem model in \textcolor{black}{the} NOMASC are replaced by JSCC-Q, while the detection process \textcolor{black}{has} also \textcolor{black}{employed} soft-to-hard mapping, which uses Euclidean distance for \textcolor{black}{the} decision. 
  \item [4)]
    DT-JSCC\cite{dtjscc}: \textcolor{black}{It consisted of a} discrete task-oriented JSCC framework \textcolor{black}{which investigates} the trade-off between the informativeness of \textcolor{black}{the} encoded data and the robustness of the distortion of received data. The encoded data is mapped into discrete representation using the Gumbel-max\cite{gumbel} method, and digital modulation is applied. We \textcolor{black}{have adopted the} DT-JSCC as a semantic codec and \textcolor{black}{modified} it for the reconstruction task. SIC and soft-to-hard mapping are employed for \textcolor{black}{the} signal detection.
  \item [5)]
    OMA Transmission: In this scheme, the transmission of the semantic information of each user is assumed to be \textcolor{black}{accomplished} in the OMA way, which means \textcolor{black}{that} there \textcolor{black}{will be} no inter-user interference. The other parts are identical to the NOMASC. The performance of this scheme can serve as an upper bound for \textcolor{black}{the} semantic transmission.
  \end{itemize}
\begin{figure*}[htbp]
  \centering
  \subfloat[Initial feature]{
        \includegraphics[height=1.5in,width=1.8in]{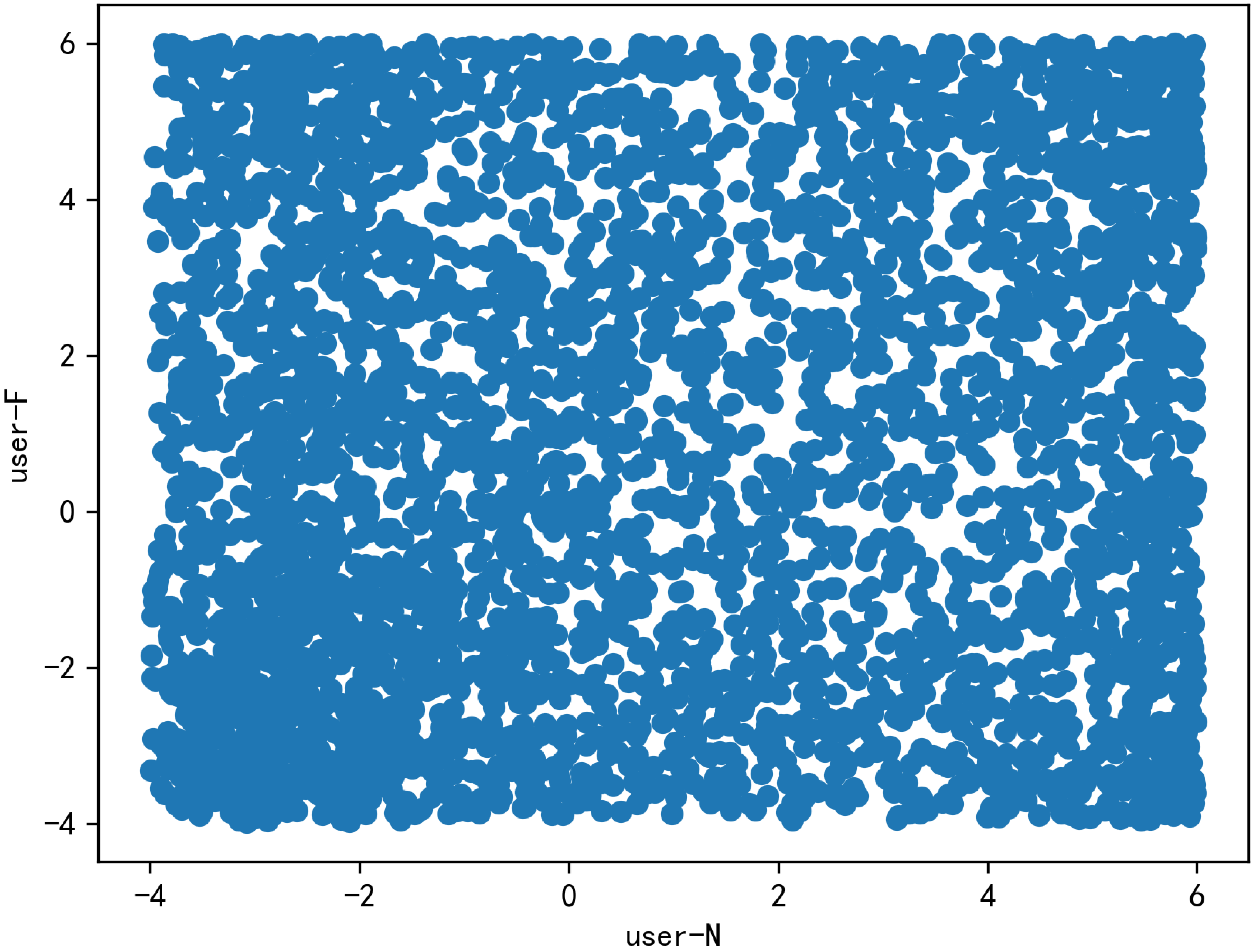}
         \label{initialfeature}}
  \quad
  \subfloat[After quantization]{
        \includegraphics[height=1.5in,width=1.8in]{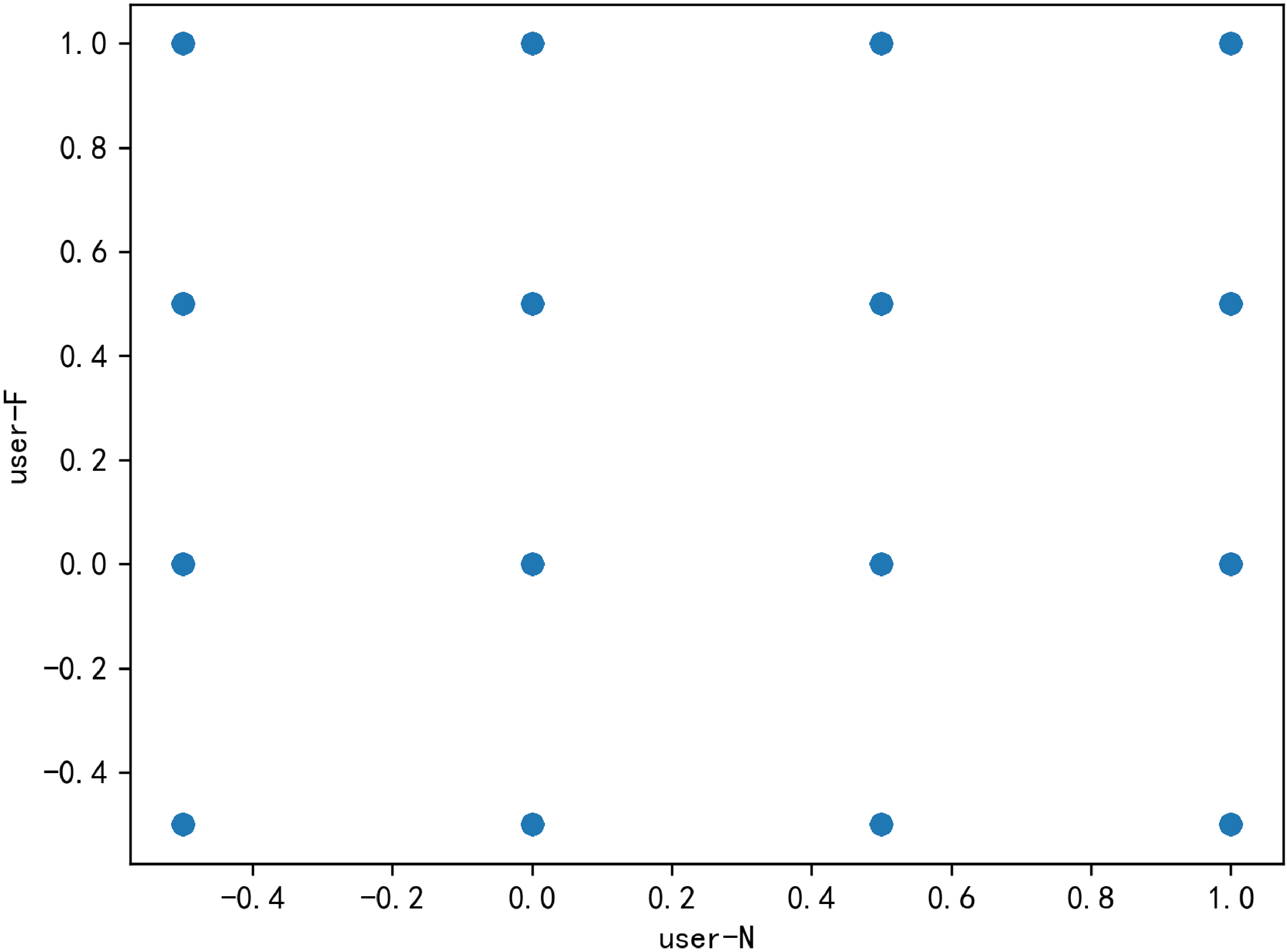}
        \label{featurequantized}}
  \quad
  \subfloat[After modulation and superimposed coding]{
        \includegraphics[height=1.5in,width=1.8in]{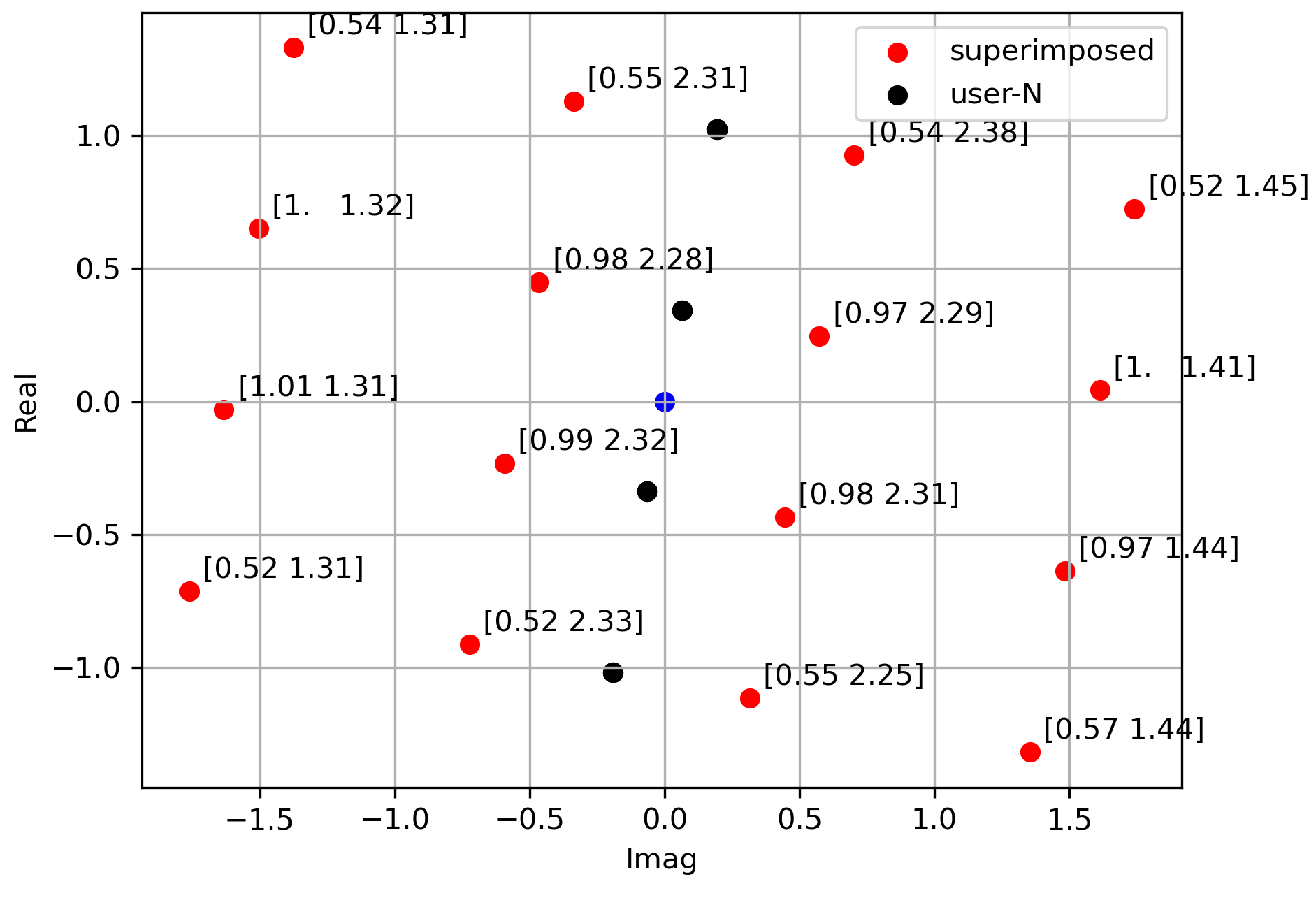}
        \label{featuremodulated}}\\
  \caption{The constellation points of the original feature, quantization set, and modulation set. The initial feature is densely distributed between [-4,6], and after \textcolor{black}{the} quantization, each element in the feature takes on a value \textcolor{black}{from} a finite constellation set. After \textcolor{black}{the} modulation, each constellation point is mapped into a two-dimensional symbol.}
  \label{featureprocess}
\end{figure*}

\subsection{Quantization and Modulation of \textcolor{black}{the Features}}
The process of asymmetric quantization and modulation is depicted in Fig. \ref{featureprocess}, where the initial output feature of the semantic encoder is continuous, and it can be observed that each element in it is distributed between $[-4,6]$. For a quantization order $m=2$, the quantization constellation set is shown in Fig. \ref{featurequantized}, where each element in the feature can only take a value from 4 potential constellations. And as analyzed in section \uppercase\expandafter{\romannumeral2}, the quantization constellation set contains zero points. The composite symbol constellation points after \textcolor{black}{the} superimposed coding are plotted in Fig. \ref{featuremodulated}. The constellation points can be seen rotating at an angle around the zero point. The constellations for user-N are also plotted, which are the center of each group of composite constellations and can be regarded as a mutation of the M-ary amplitude shift keying (MASK) constellation. The MSE between the original quantized feature and the restored one is printed next to each constellation point. It can be \textcolor{black}{generalized} that the MSE of the message carried in the constellations located near \textcolor{black}{the} zero point is higher, \textcolor{black}{whilst} for the constellations which are more distant from \textcolor{black}{the} zero point, the MSE is lower since the effect of noise on the constellation with higher amplitude is weaker. As a result, the model can learn to give unequal protection to the source by adjusting the weight parameter in the model.

\begin{figure*}[htbp]
  \centering
  \subfloat[Cifar\&Cifar case under AWGN channel]{
        \includegraphics[height=1.3in,width=3in]{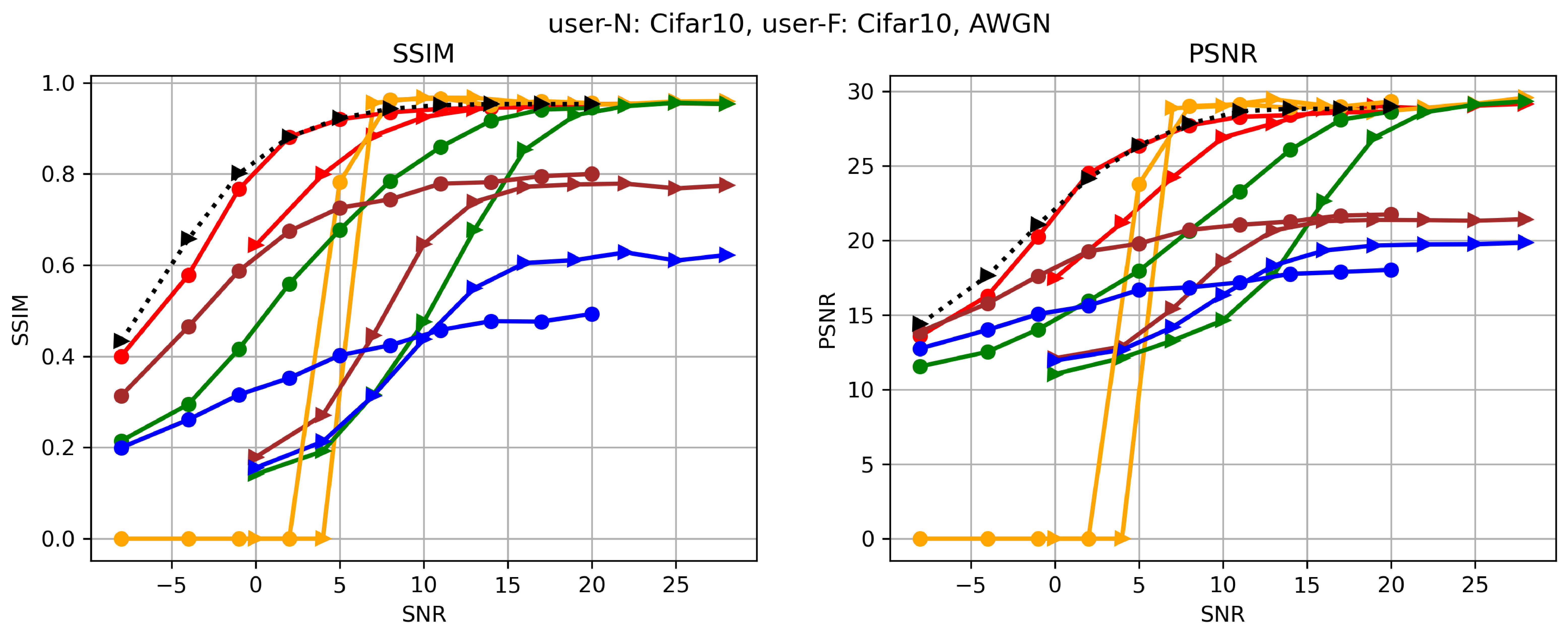}
         \label{cifarcifarawgn}}
  \quad
  \subfloat[Cifar\&MNIST case under AWGN channel]{
        \includegraphics[height=1.3in,width=3in]{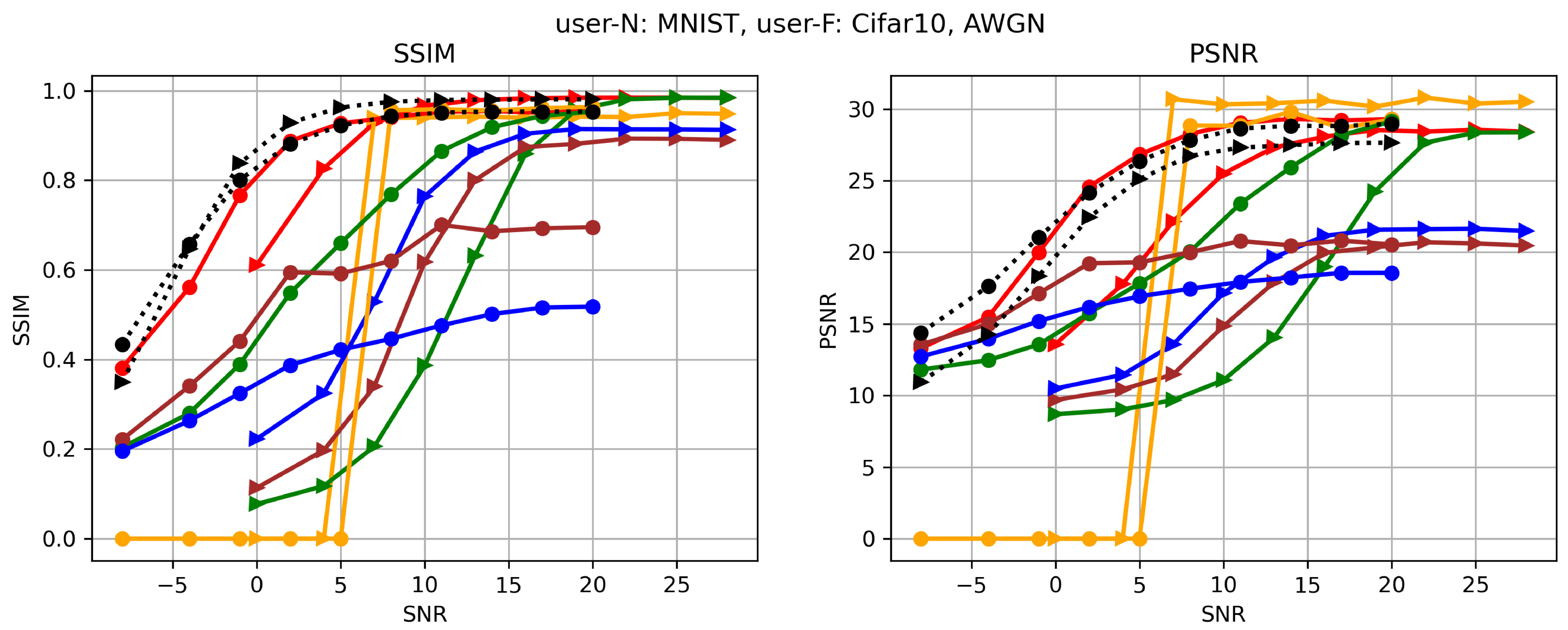}
        \label{cifarmnistawgn}}
  \quad
  \subfloat[Cifar\&Europarl case under AWGN channel]{
        \includegraphics[height=1.5in,width=4.8in]{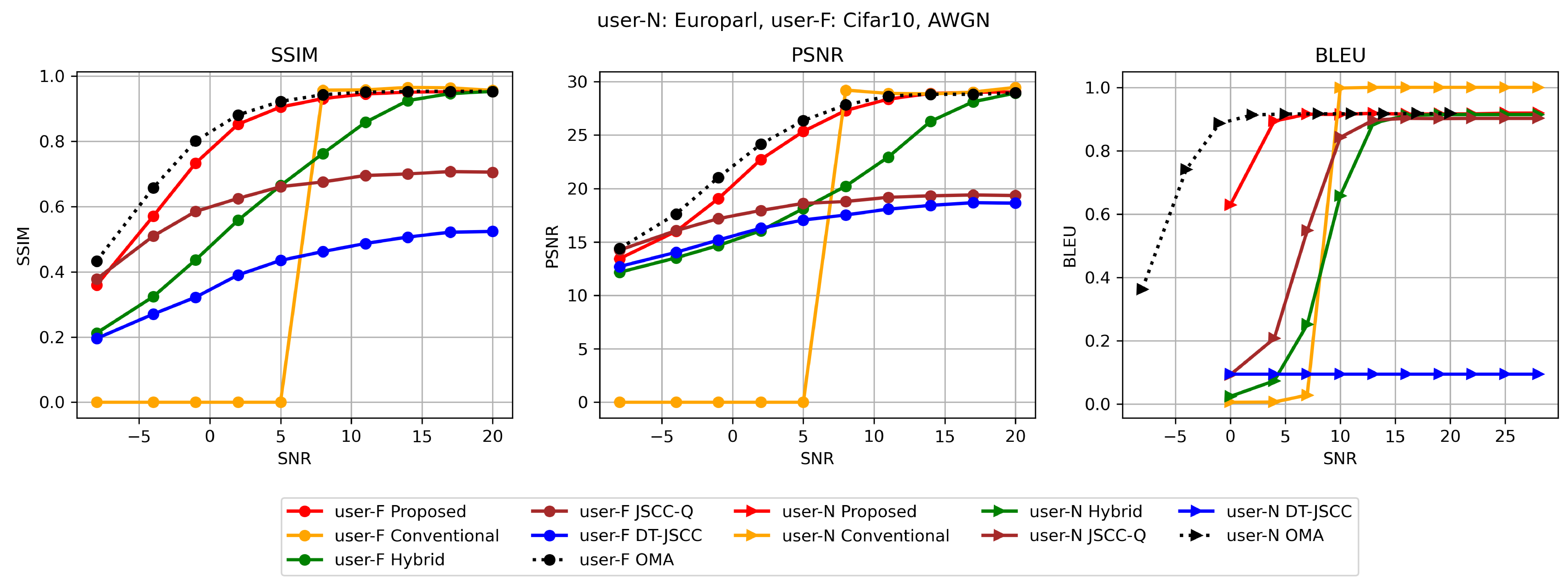}
        \label{cifareuroparlawgn}}\\
  \caption{Transmission performance under the AWGN channel, where (a) is the Cifar\&Cifar case, (b) is the Cifar\&MNIST case, and (c) is the Cifar\&Europarl case. For image semantic transmission, SSIM and PSNR are used for evaluation, whereas BLEU (1-gram) is used for text semantic transmission.}
  \label{awgn}
\end{figure*}
\begin{figure*}[htbp]
  \centering
  \subfloat[Cifar\&Cifar case under Rayleigh fading channel]{
        \includegraphics[height=1.2in,width=3in]{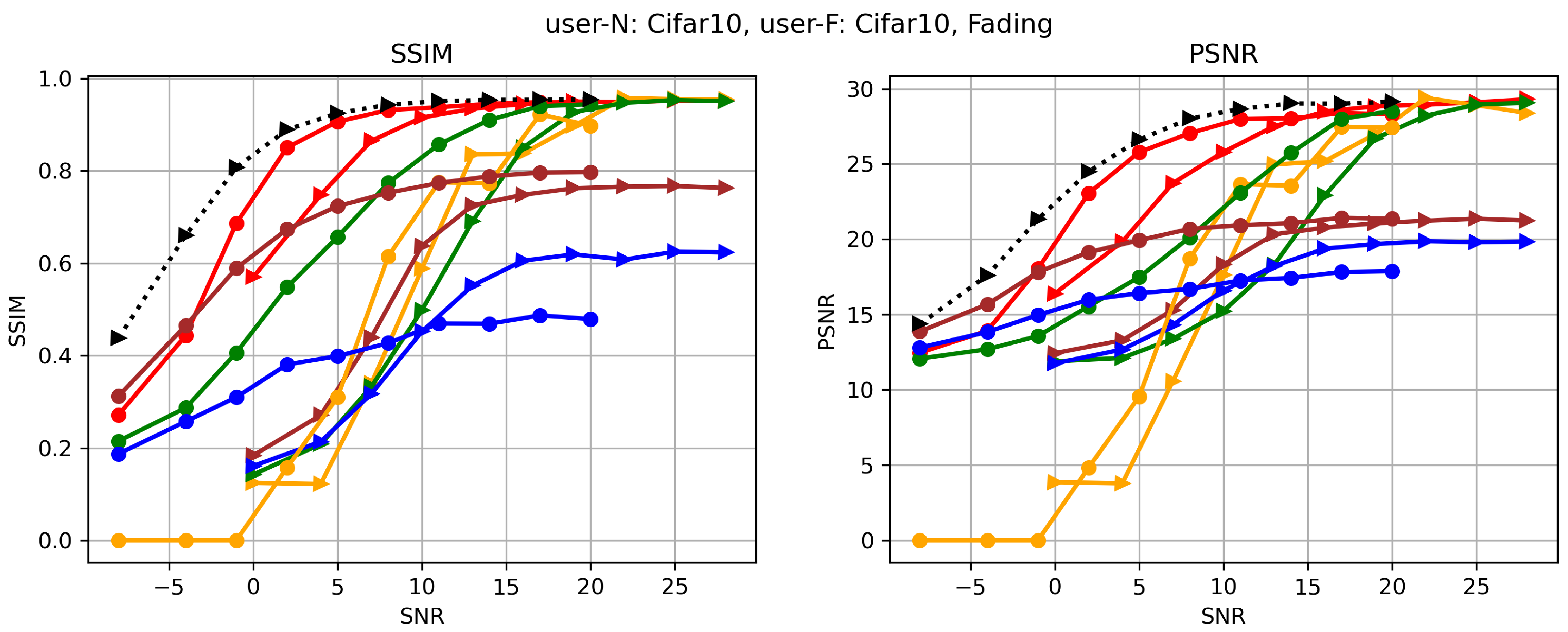}
         \label{cifarcifarrayleigh}}
  \quad
  \subfloat[Cifar\&MNIST case under Rayleigh fading channel]{
        \includegraphics[height=1.2in,width=3in]{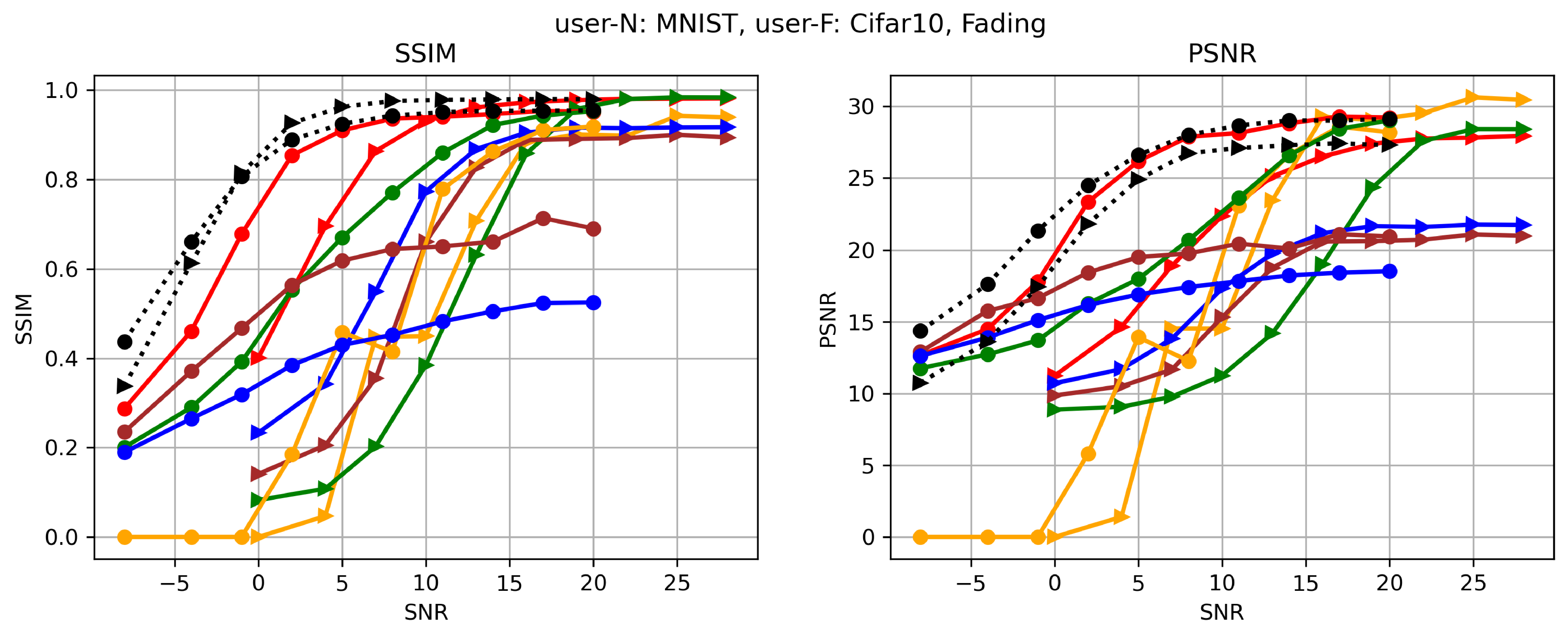}
        \label{cifarmnistrayleigh}}
  \quad
  \subfloat[Cifar\&Europarl case under Rayleigh fading channel]{
        \includegraphics[height=1.5in,width=4.8in]{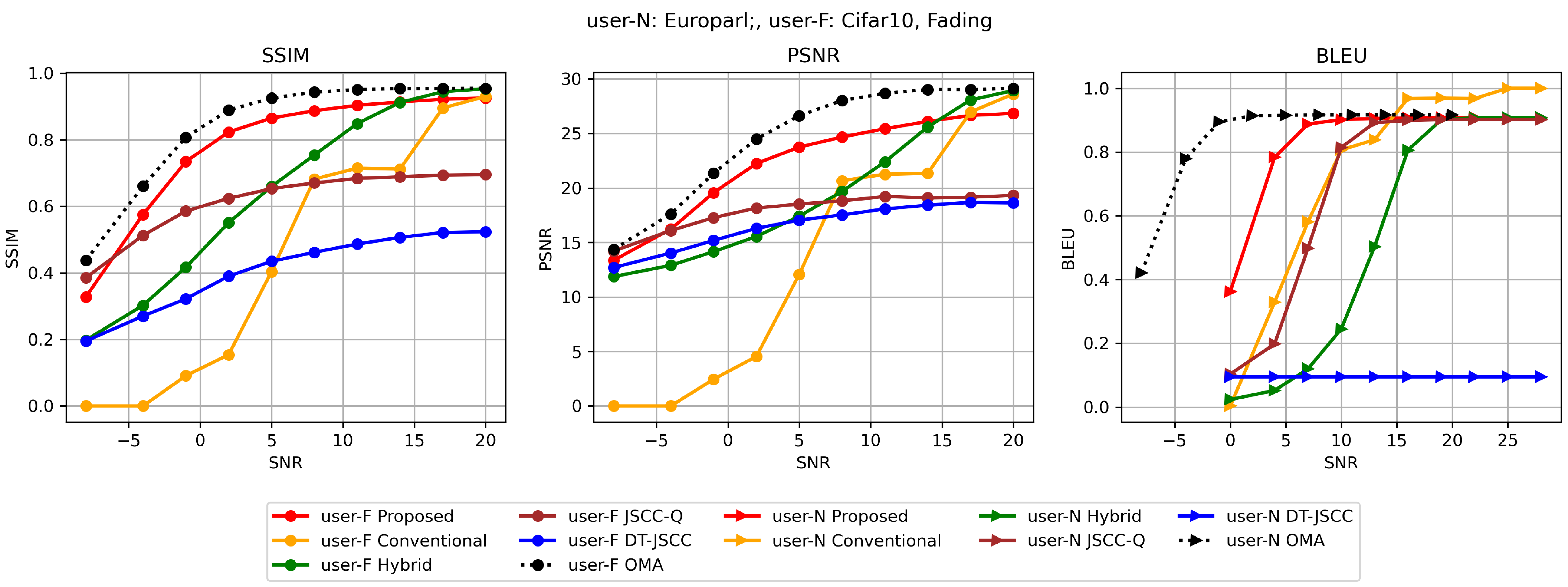}
        \label{cifareuroparlrayleigh}}\\
  \caption{Transmission performance under the Rayleigh fading channel, the simulation setting is the same as the AWGN case except for the channel type.}
  \label{rayleigh}
\end{figure*}
\subsection{Transmission Performance}
The performance of \textcolor{black}{the} NOMASC is evaluated alongside a series of baseline approaches under AWGN and fading channels. \textcolor{black}{As} shown in Figs. \ref{awgn} and \ref{rayleigh}, the proposed NOMASC scheme \textcolor{black}{attains} good performance under low and high SNR, outperforming all other baseline methods. Additionally, its performance is almost as good as the OMA scenario, which means the demodulator model can almost thoroughly eliminate \textcolor{black}{the} inter-user interference. The conventional method \textcolor{black}{also attains} excellent performance under high SNR. However, it \textcolor{black}{has} suddenly \textcolor{black}{deteriorated} when the channel condition is bad, which is called the “cliff effect”. It \textcolor{black}{may} be noted that \textcolor{black}{none of the} learning-based methods \textcolor{black}{has shown} the “cliff effect” \textcolor{black}{but} a gentler decline. The hybrid method performs well under high SNR, but \textcolor{black}{not as good as} the proposed method under low SNR, which \textcolor{black}{suggests a poorer} ability to protect \textcolor{black}{the} information from channel noise. And this has also verified the effectiveness of the modem model. For the JSCC-Q and DT-JSCC methods, \textcolor{black}{their} performances are not bad when processing images with a very small resolution like MNIST, but when it comes to larger images, their performances are much \textcolor{black}{poorer} than NOMASC, \textcolor{black}{and} the reason behind this could be the difficulty of signal detection. The gradient approximation used in the soft-to-hard decision process might hinder backpropagation \textcolor{black}{and effect the accuracy of detection}. The problem \textcolor{black}{become} even \textcolor{black}{worse} in the \textcolor{black}{case of} Europarl\&CIFAR, \textcolor{black}{as} it can be \textcolor{black}{noted} from Fig. \ref{cifareuroparlawgn} that the service quality for text user has broken down to a flat line.

\textcolor{black}{During} the simulation, it is assumed that the type of fading channel is Rayleigh, where the channel coefficients are generated following $\mathbf{h} \sim \mathcal{CN}(0, 1)$. On the receiver side, the CSI of the downlink channel is assumed to be known, so that equalization can be performed. It can be observed from Fig. \ref{rayleigh} that the performance of each method is similar to the AWGN case, where \textcolor{black}{the} NOMASC yields the best results. Under both AWGN and fading channels, the conventional method \textcolor{black}{has outperformed} all learning-based methods at high SNR, which is inevitable given that the learning-based method may be seen as a maximum likelihood estimator of the original data. The approximated float numbers may contain mistakes as long as noise and interference remain.
\begin{figure}[h]
\centering
\includegraphics[width=3.5in]{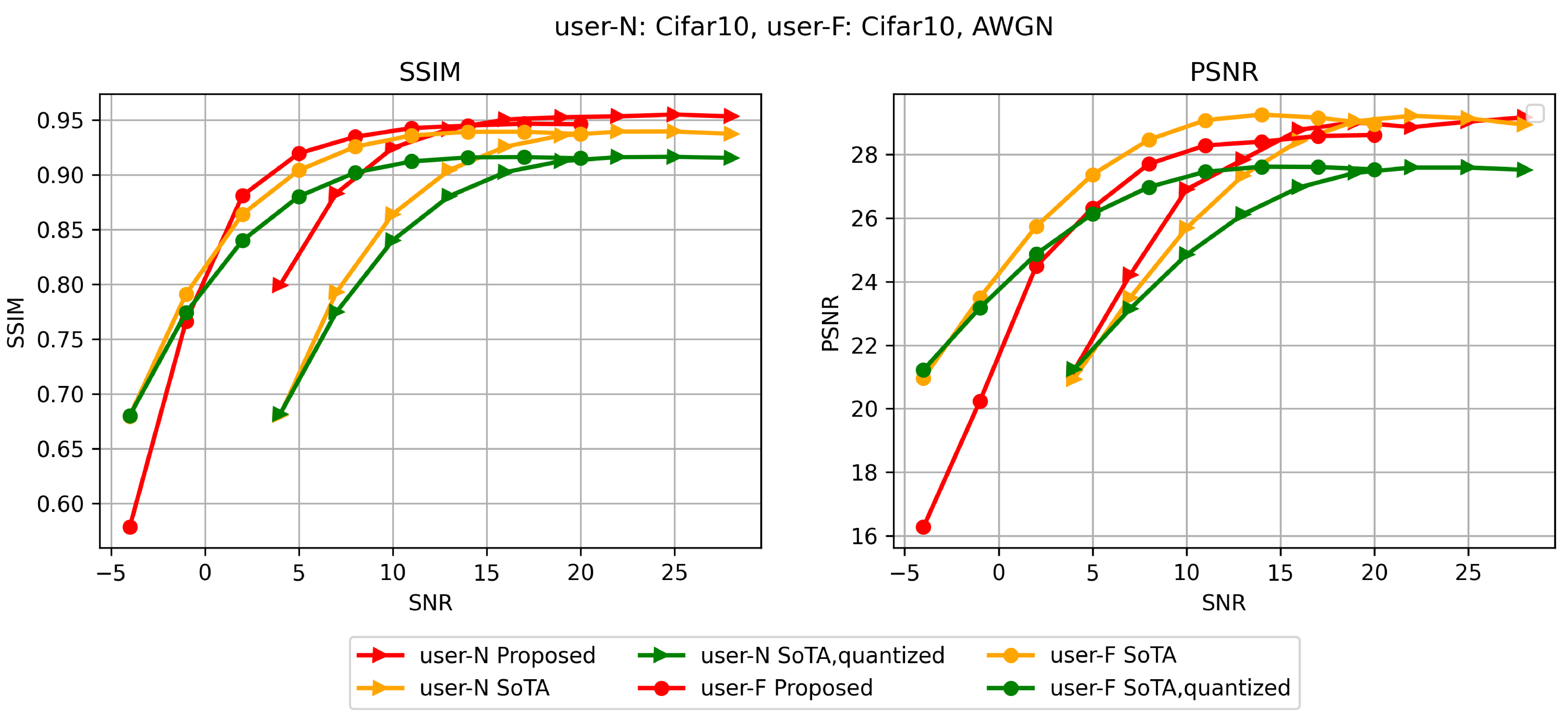}
\caption{Comparison between NOMASC and the state-of-the-art scheme }
\label{sotacompare}
\end{figure}

\subsection{\textcolor{black}{Comparsion with the State-of-the-art Scheme}}
\textcolor{black}{The state-of-the-art NOMA semantic scheme, as proposed in \cite{nomajscc}, has employed the curriculum learning technique along with the attention feature module for training the end-to-end image transmission framework. Simulation results have validated its advantage over the TDMA scheme and effectiveness under various channel conditions. Since it can only serve users with the same image dataset, we make comparison with it separately in this subsection. Although this work was conducted within the context of multiple access channel, we have made modification for comparing it with our NOMASC under the same broadcast channel setting.}

\textcolor{black}{We have mainly compared the NOMASC with the full-resolution state-of-the-art (SoTA) scheme and the one with feature quantization operation for a fair comparison (same bit per pixel). It may be noted from Fig. \ref{sotacompare} that the main advantage of the NOMASC is on user-N, which validates the effectiveness of the modulation model. On the other hand, the SoTA scheme has the advantage in terms of PSNR, which means that it can produce reconstructed images with lower MSE. Furthermore, SoTA scheme attains better performance under low SNR, which can be attributed to the adaptive learning ability of the attention feature module. }
\begin{figure*}[htbp]
	\centering
	\subfloat[Robustness to channel estimation error]{
	\label{estimationerror}
		\centering
		\includegraphics[height=1.3in,width=3in]{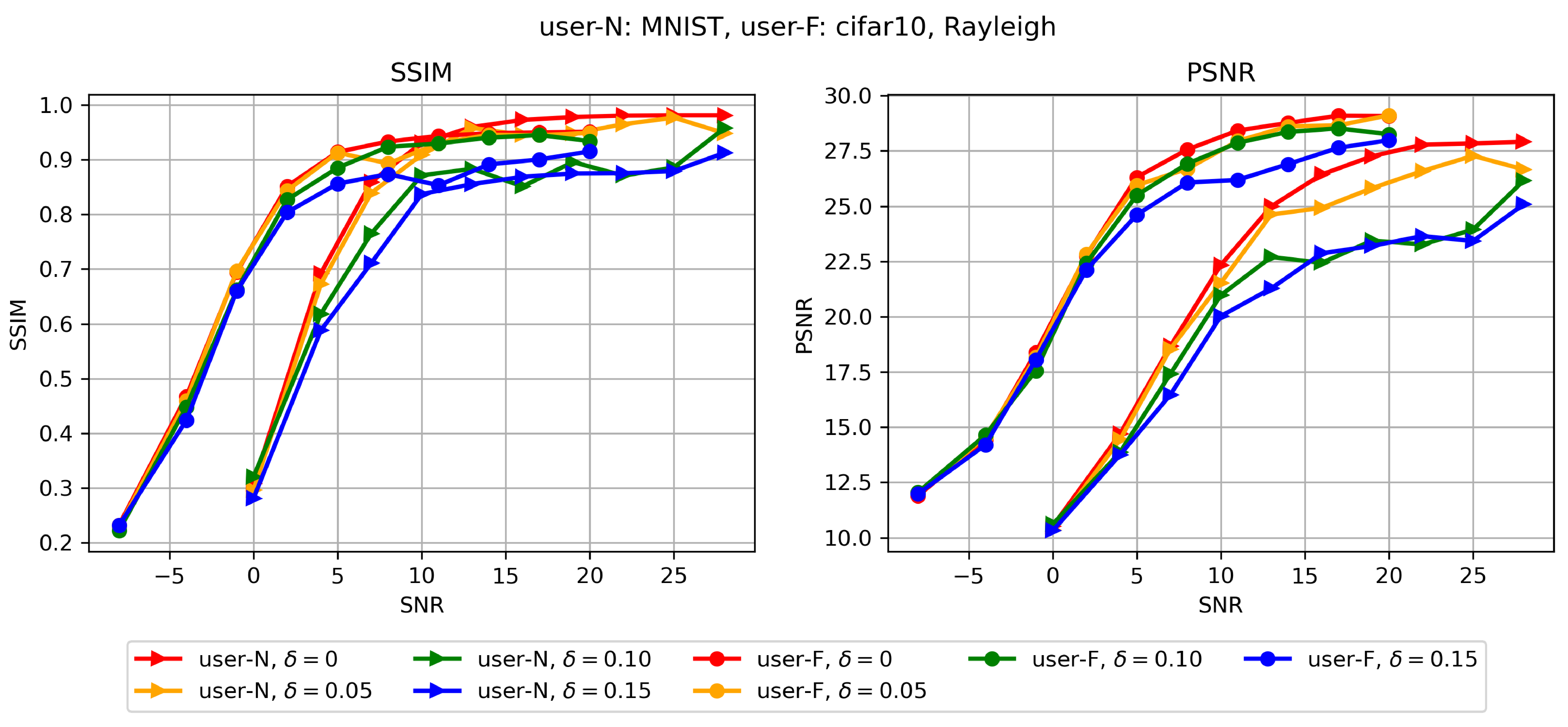}
	}
	\quad
	\subfloat[Robustness to different power allocation scheme]{
	\label{powerallocation}
		\centering
		\includegraphics[height=1.3in,width=3in]{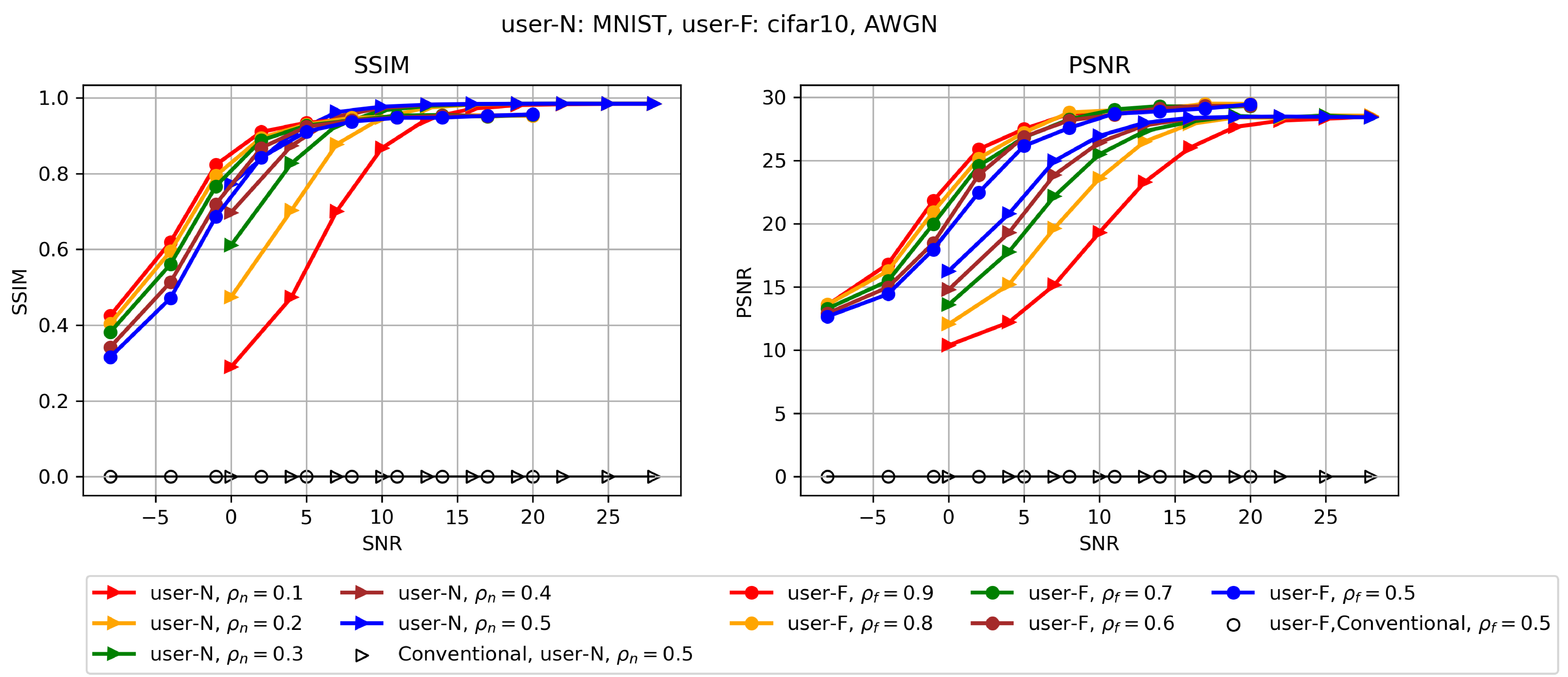}
	}
	\quad
 	\subfloat[Effect of different modulation order]{
 	\label{modulationorder}
		\centering
		\includegraphics[height=1.3in,width=3in]{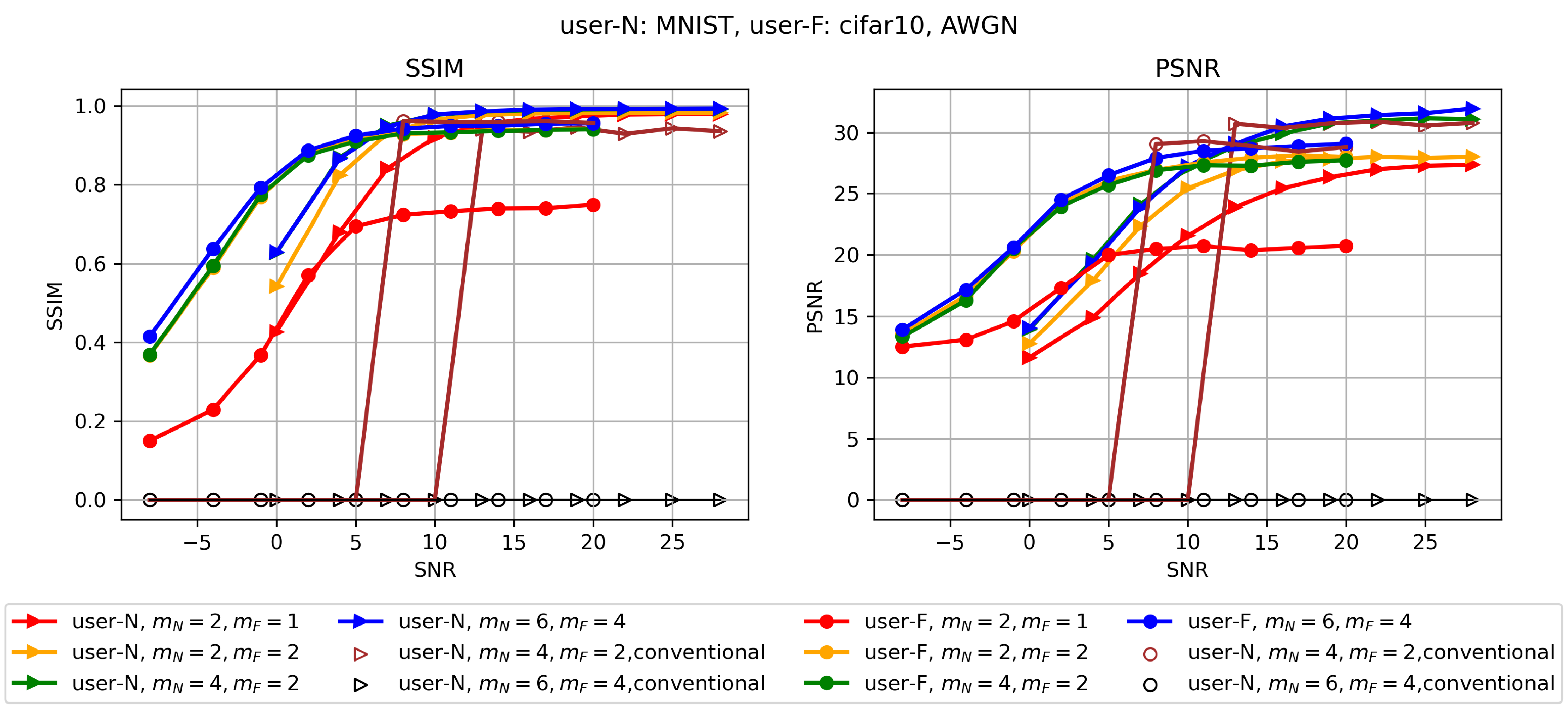}
	}
 	\quad
 	 	\subfloat[Comparison to F-JSCC scheme in mismatched scenario]{
 	 	\label{FJSCC}
		\centering
		\includegraphics[height=1.3in,width=3in]{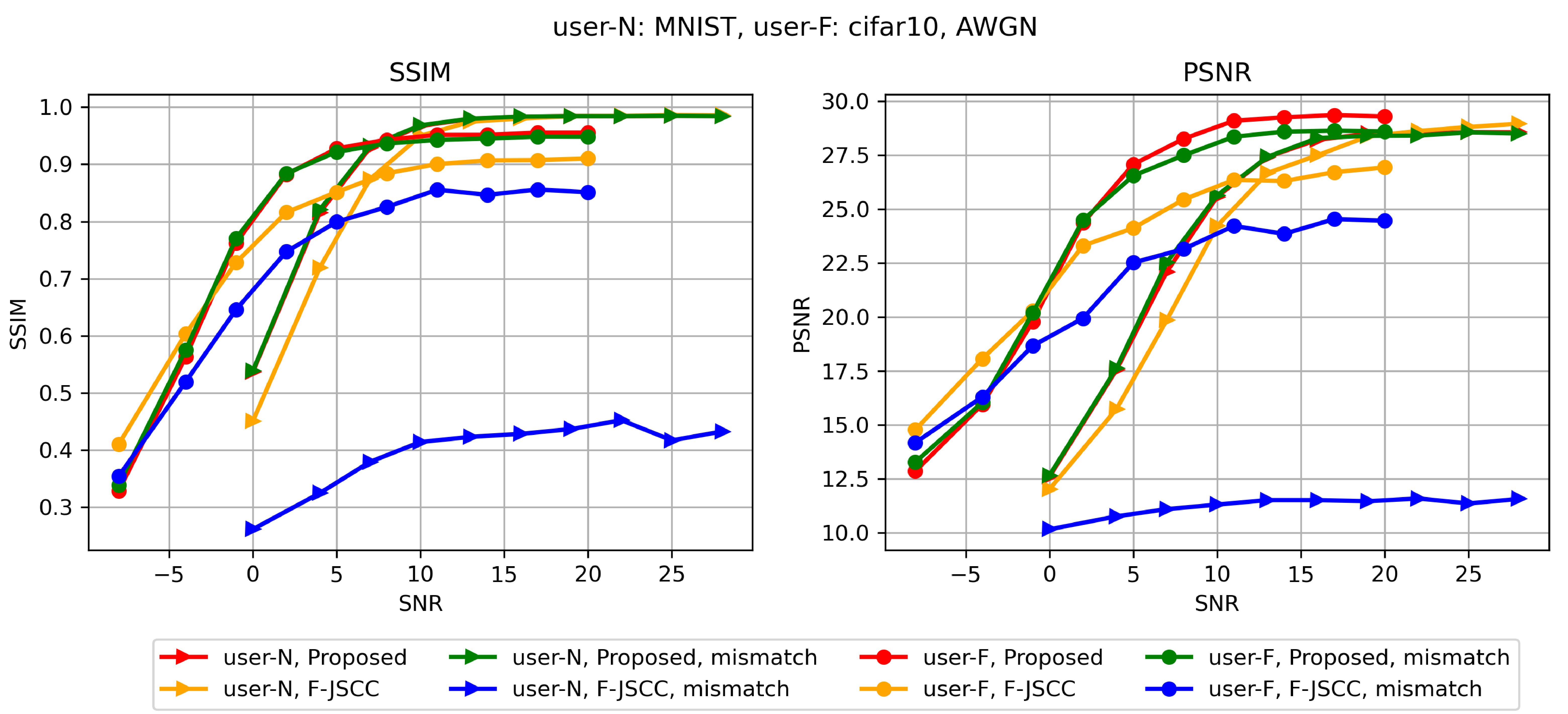}
	}
 	\centering
	\caption{Robustness of the model to a variety of conditions.}
	\label{robustness}
\end{figure*}

\subsection{Robustness Evaluation}
Models in the system are deployed to \textcolor{black}{the} end devices and BS after \textcolor{black}{the} training. Since the parameter settings used in the training process can be significantly different from the practical scenarios, even though the model can be fine-tuned using a short period for adaptation, it may cause a non-negligible delay and affect the whole system's efficiency. Investigating the proposed system's robustness in various test conditions is therefore important.

\subsubsection{Effect of Channel Estimation Error}
In the practical environment, channel estimation cannot be perfect, and the channel estimation error can be modeled as follows:
\begin{align}\label{deqn_ex1a}
\hat{\mathbf{h}} = \mathbf{h} + \mathbf{\delta}*\mathbf{e},
\end{align}
where $\mathbf{h}$ is the actual channel coefficient and $\hat{\mathbf{h}}$ is the estimated channel coefficient, $\mathbf{e}$ is the estimated error, and the degree of error is controlled by $\mathbf{\delta}$. The simulation result is shown in Fig. \ref{estimationerror}. It can be seen that the performance of \textcolor{black}{the} NOMASC rapidly \textcolor{black}{deteriorates} as the error factor $\mathbf{\delta}$ increases. When $\mathbf{\delta}$ is less than 0.1, such performance degradation is not notable, and in the worst case, where $\mathbf{\delta} = 0.15$, the SSIM and PSNR decline by roughly 0.1 and 4 dB, respectively.
\subsubsection{Effect of Power Allocation Scheme}
Power allocation for each user in the group needs to be calculated based on the CSI feedback. Since the model is trained using a predetermined power allocation scheme, \textcolor{black}{the} BS may need to recompute the power allocation after the CSI of one user's link has changed, or it may have an impact on the transmission performance. The robustness of the model to the variety of power allocation schemes is shown in Fig. \ref{powerallocation}. The service quality of user-F is not very sensitive to the variety of power allocation factors, and the curves are close to one another. For user-N, it is more sensitive since $\rho_{N}$ is smaller than $\rho_{F}$. Conventional methods are also used for comparison, in the equal power allocation situation where $\rho_{N}=\rho_{F}=0.5$, the performance of the conventional method \textcolor{black}{has broken} down completely where SSIM and PSNR are \textcolor{black}{both} zeros, which indicates that none of the text images can be successfully reconstructed. This is because when the power allocation factors are equal, traditional SIC detection \textcolor{black}{may fail} to demodulate either of the user signals. The learning-based approach, however, can produce meaningful results.
\subsubsection{Effect of Different Modulation Order}
For conventional methods, a high modulation order can enhance the transmission efficiency \textcolor{black}{whilst} increase the difficulty of signal detection. For the proposed NOMASC, modulation order is relevant to the size of the quantization set, which also has an impact on how well the end-to-end semantic codec performs. The quantization error for each component of the feature can be reduced with additional quantization points \textcolor{black}{to preserve} more information. The simulation result is shown in Fig. \ref{modulationorder}. The lowest transmission performance occurs when the modulation order is 2, as there are only two options for quantization, causing unavoidable significant information loss and quantization error. The transmission performance improves with higher quantization orders. For the conventional method, the performance is still good under $m_{N}=4, m_{F}=2$, \textcolor{black}{but has failed} to recover any meaningful information at the receiver under $m_{N}=6, m_{F}=4$. \textcolor{black}{By contrast}, it can be seen that the issue of detection difficulty does not show up even in the highest modulation order, $m_{N}=6, m_{F}=4$, which is also an advantage of the proposed scheme compared \textcolor{black}{with} the conventional method.
\subsubsection{Comparison \textcolor{black}{with the} F-JSCC Scheme}
In this section, we have compared \textcolor{black}{the} NOMASC \textcolor{black}{with} the full-resolution JSCC (F-JSCC) scheme, where the channel input is not constrained and the feature generated by the semantic encoder does not \textcolor{black}{undergo} the quantization process. However, the generalization ability of this method is \textcolor{black}{not as good as the} NOMASC since the test environment \textcolor{black}{is required} to be consistent with the training environment. The decoder can only estimate the desired data accurately \textcolor{black}{should} the received signal follow a specific distribution. For example, if the end-to-end F-JSCC codec is trained under the Cifar\&Cifar scenario, the decoder can only deal with the superimposition of two encoded Cifar images. Unfortunately, it appears that in a real-world setting, this \textcolor{black}{may} not always be satisfied. Using \textcolor{black}{the} NOMASC, the demodulator model is capable of dealing with any two superimposed semantic signals as long as the quantization orders of them are the same. This is verified by simulation, and the result is shown in Fig. \ref{FJSCC} where the F-JSCC model \textcolor{black}{attains} a close performance to \textcolor{black}{the} NOMASC when the test environment is identical to the training phase (MNIST\&Cifar). Additionally, during the mismatched \textcolor{black}{testing} procedure, the models of both F-JSCC and NOMASC for user-F are trained using a single Cifar10 dataset, and the models for user-N are trained using the MNIST dataset. It \textcolor{black}{may be noted} that when testing both models under the MNIST\&CIFAR scenario, the performance of user-N in the F-JSCC scheme \textcolor{black}{has substantially declined}, whereas \textcolor{black}{the} NOMASC continues to perform well. This \textcolor{black}{is in keeping} with our analysis that the decoder of user-N \textcolor{black}{has difficulty} to deal with the unfamiliar superimposed signals.
\subsection{\textcolor{black}{Capability of Serving Three Users}}
\begin{figure*}[htbp]
	\centering
	\subfloat[\textcolor{black}{Superimposed constellation points in the three-user case}]{
		\centering
		\includegraphics[height=1.5in,width=2in]{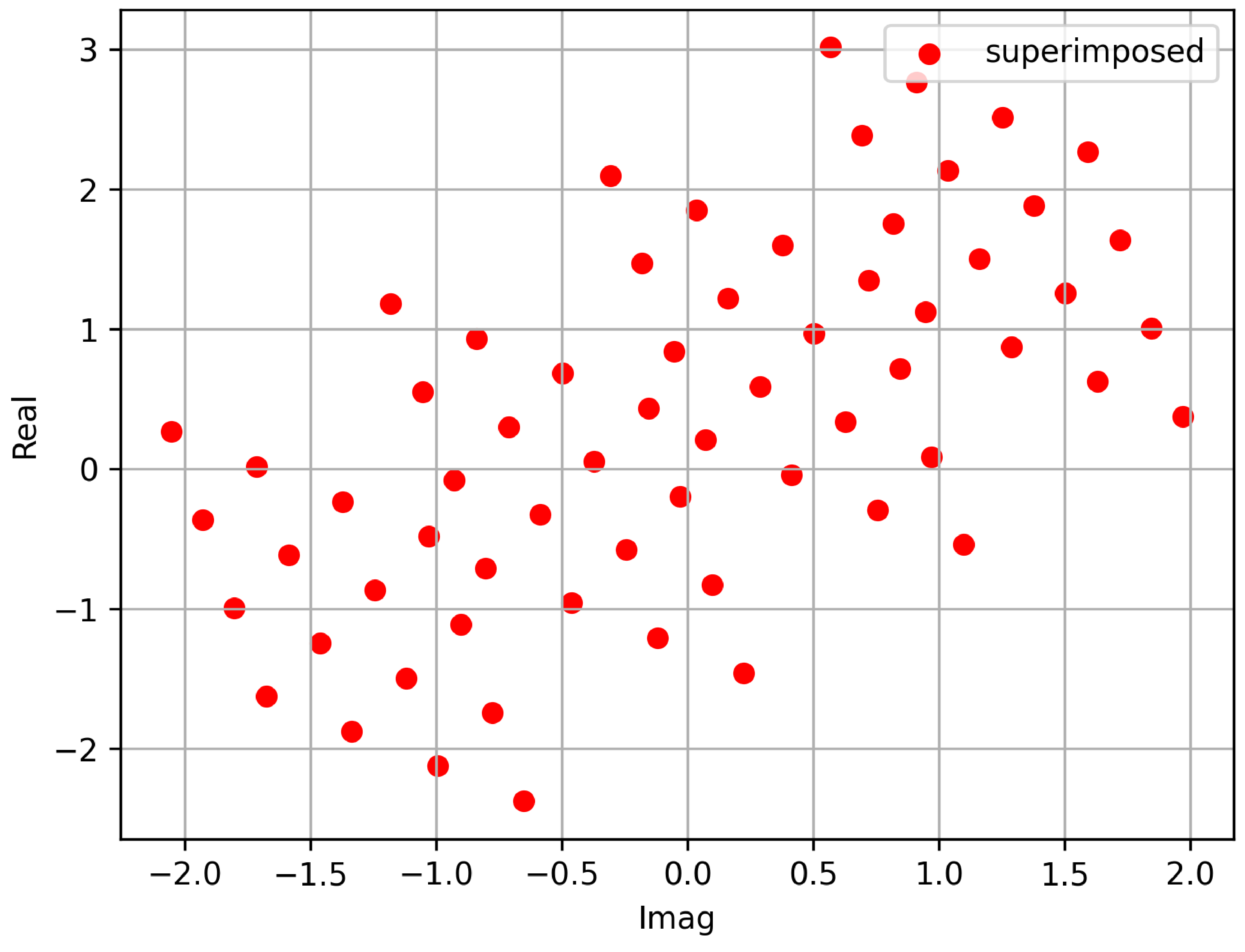}
		\label{3user_constellations}
	}
	\quad
	\subfloat[\textcolor{black}{SSIM evaluation result in the three-user case}]{
		\centering
		\includegraphics[height=1.5in,width=2in]{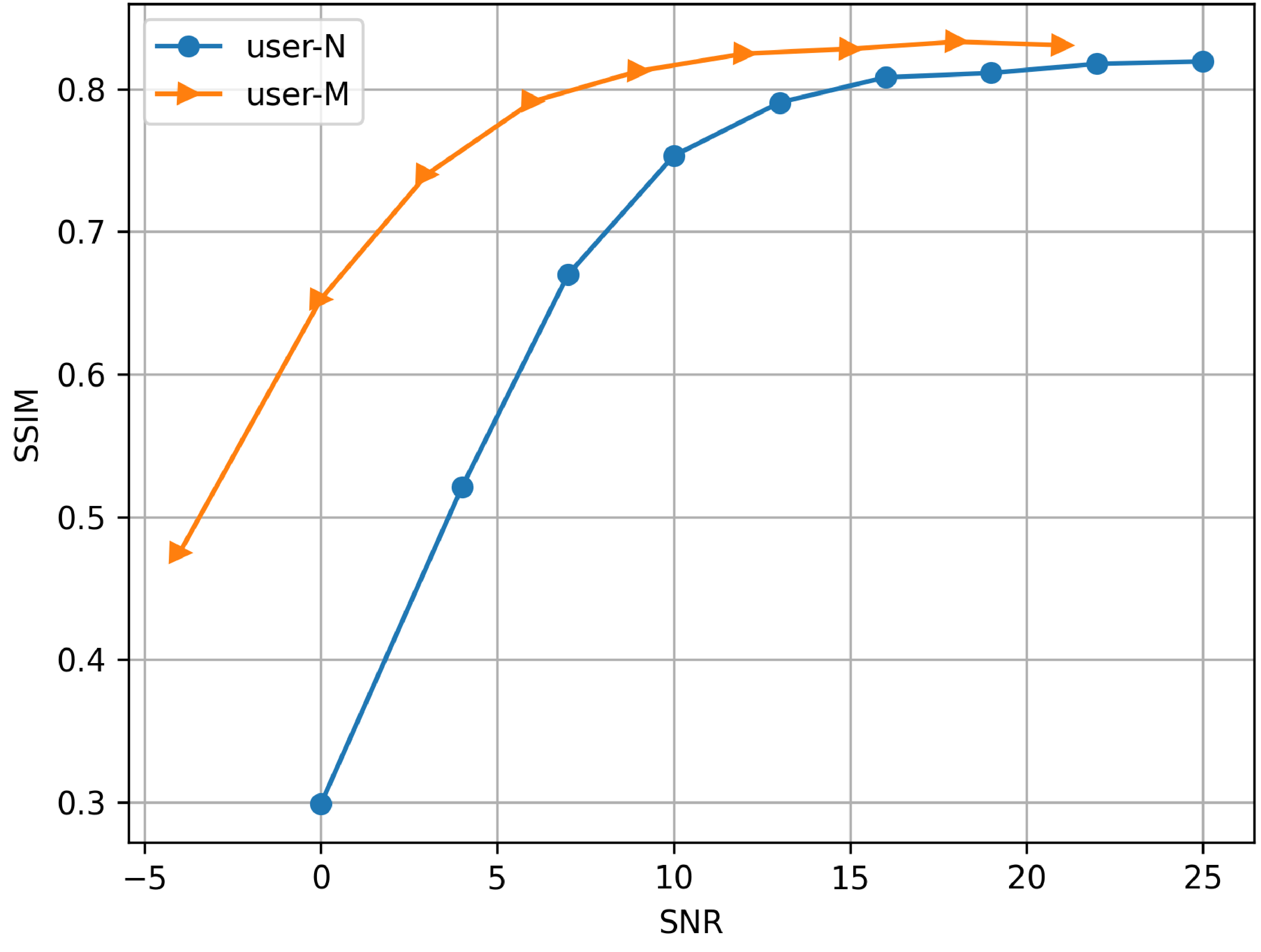}
		\label{3user_SSIM}
	}
        \quad
        \subfloat[\textcolor{black}{Examples of image transmission}]{
		\centering
		\includegraphics[height=1.75in,width=1.75in]{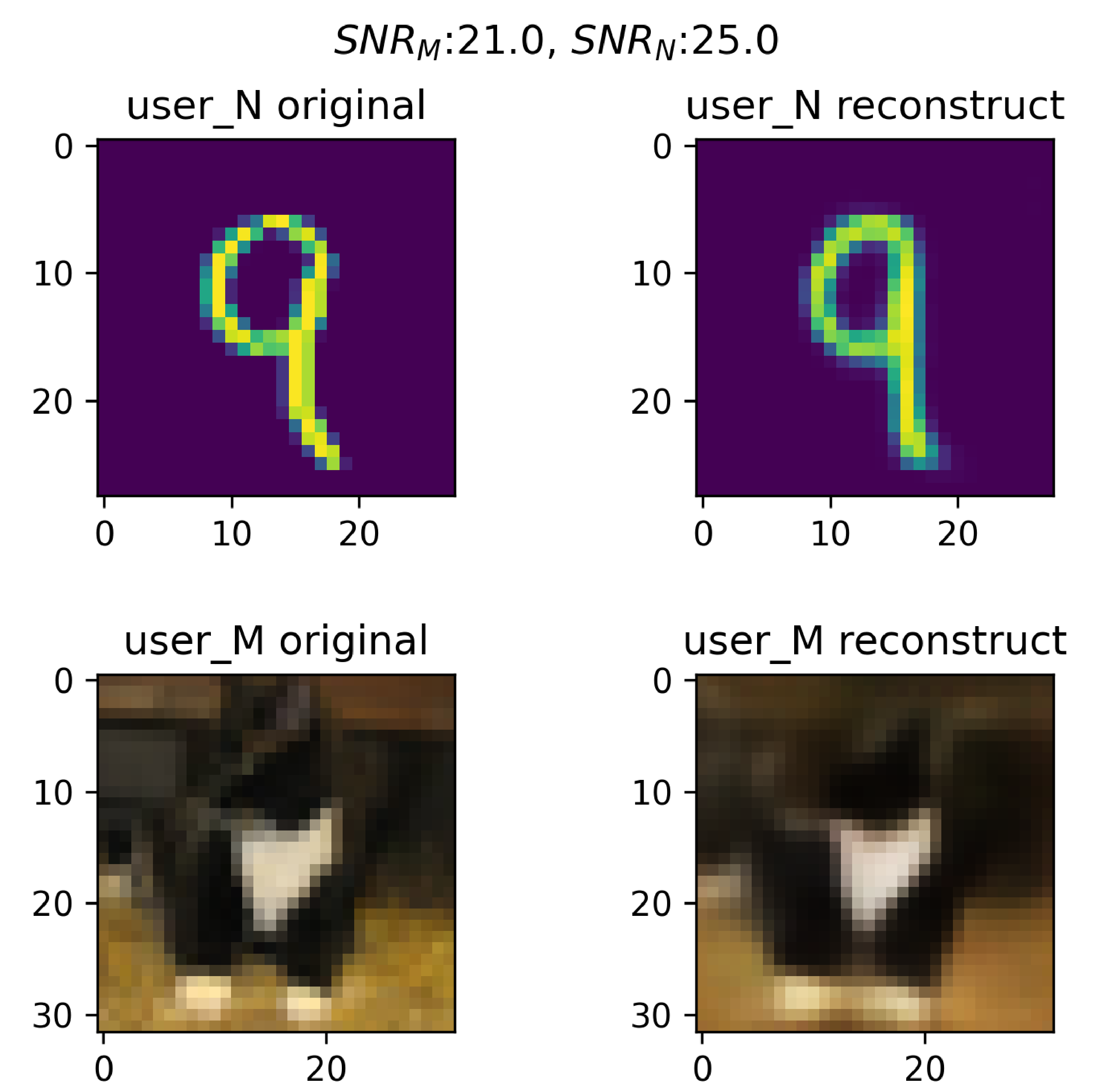}
		\label{image example}
	}
        \\
 	\subfloat[\textcolor{black}{PSNR evaluation result in the three-user case}]{
		\centering
		\includegraphics[height=1.5in,width=2in]{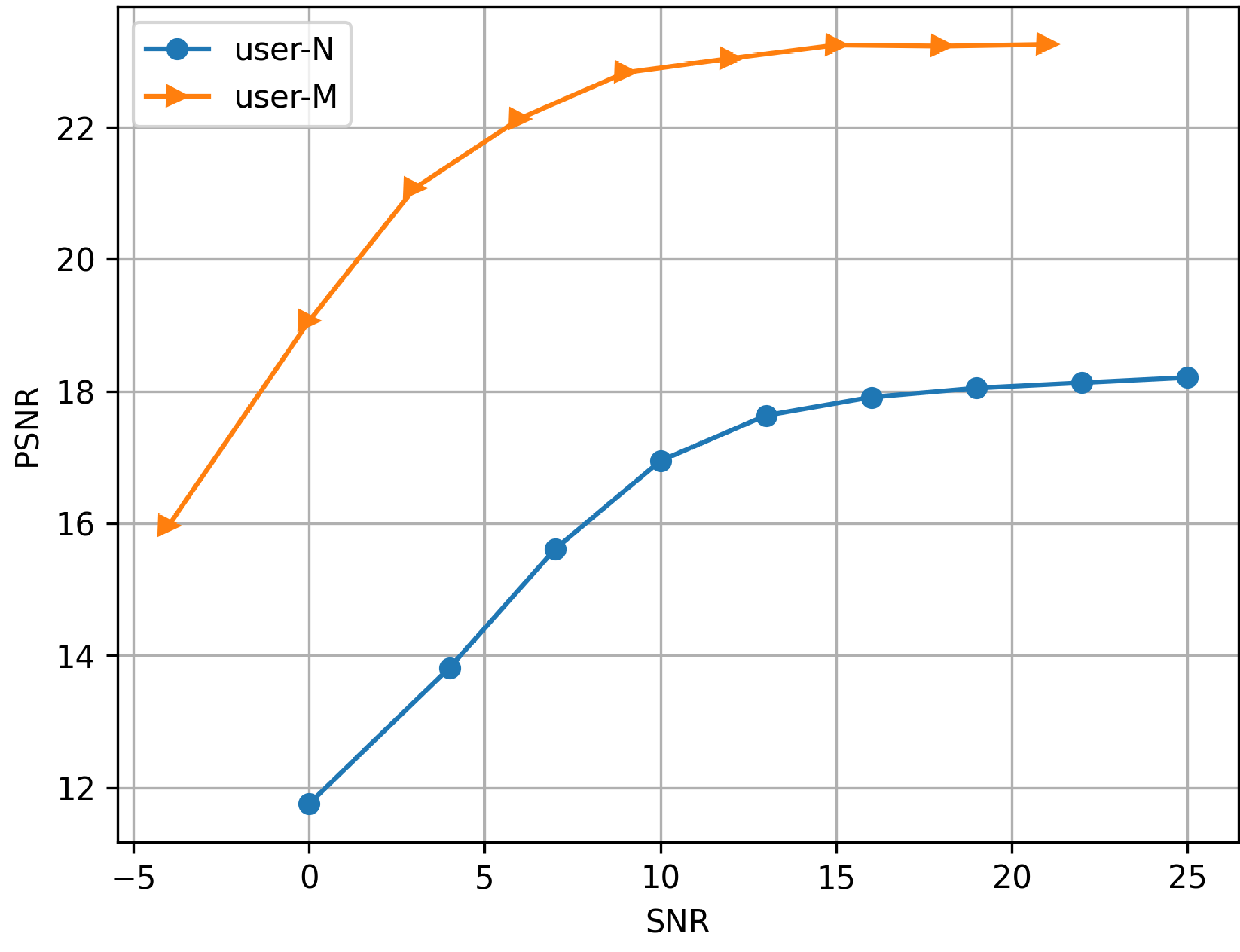}
		\label{3user_PSNR}
	}
 	\quad
 	 	\subfloat[\textcolor{black}{Sentence similarity evaluation result in the three-user case}]{
		\centering
		\includegraphics[height=1.5in,width=2in]{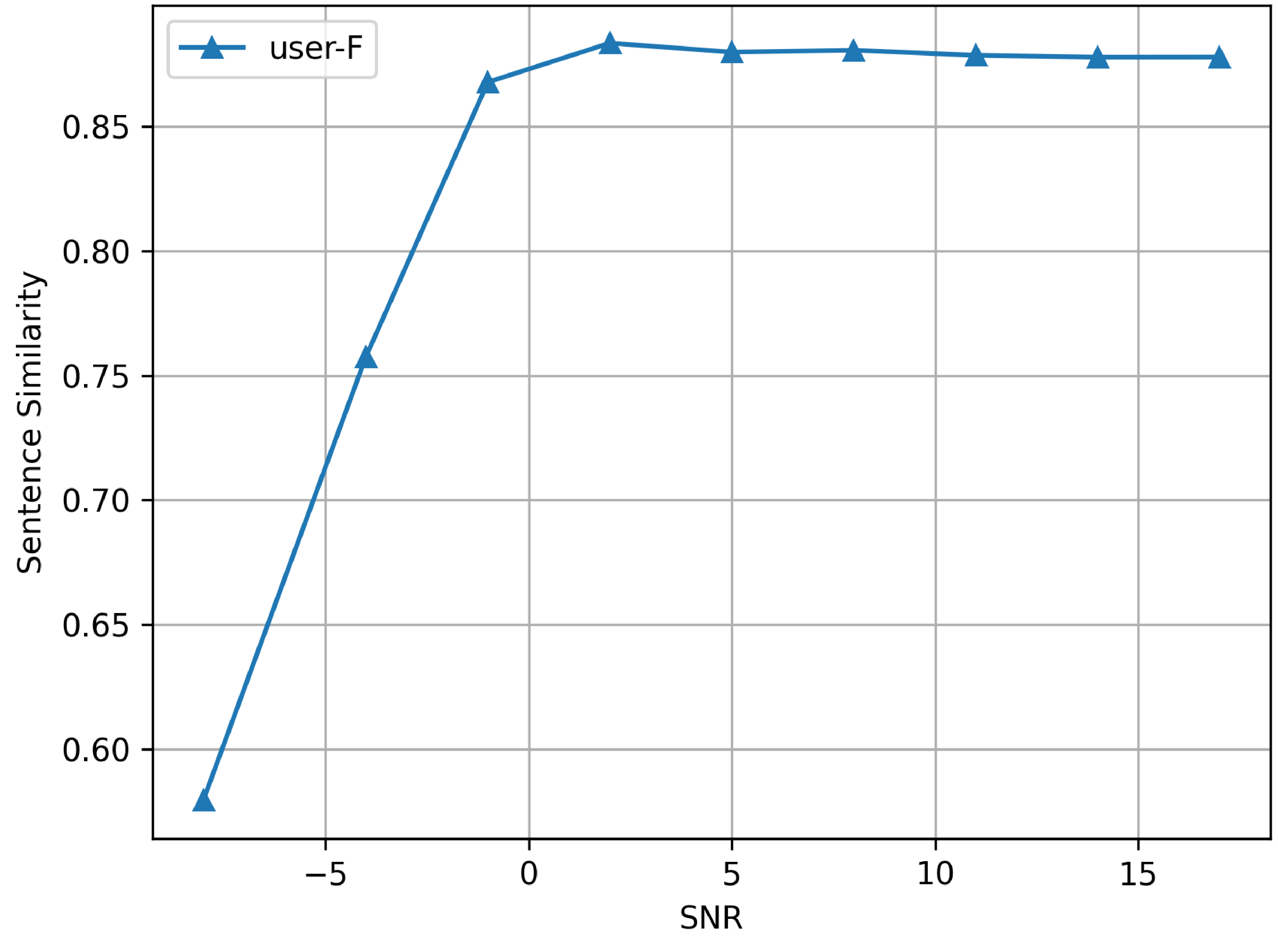}
		\label{3user_SS}
	}
        \quad
        \subfloat[\textcolor{black}{Example of sentence transmission}]{
		\centering
		\includegraphics[height=1.5in,width=2in]{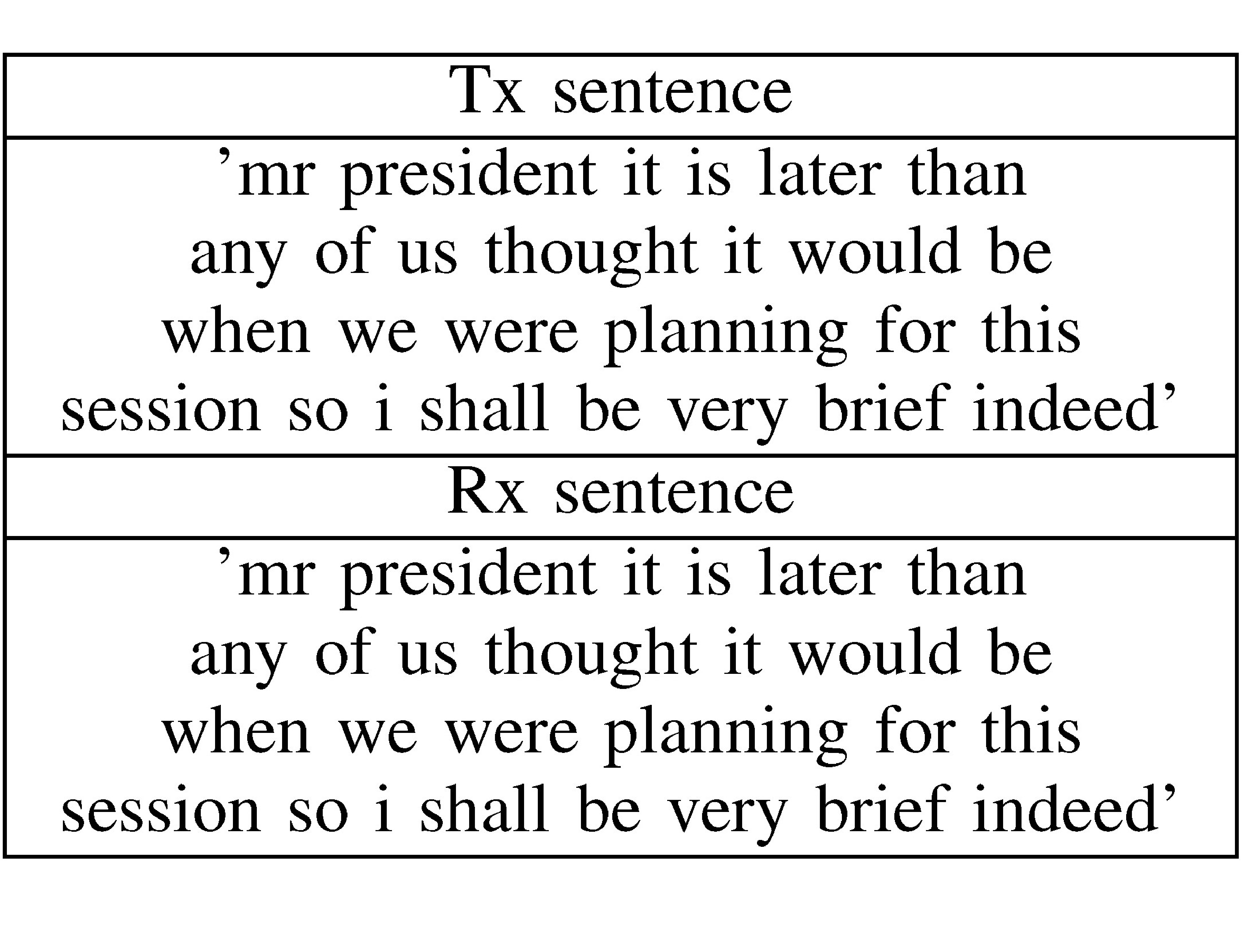}
		\label{3user_example_sentence}
	}
 	\centering
	\caption{\textcolor{black}{The simulation results of the three-user NOMASC system. The superimposed constellation points are given in (a). End-to-end reconstruction accuracy evaluation results are given in (b), (d) and (e). Examples of image and sentence transmission are given in (c) and (f).}}
	\label{three-user case}
\end{figure*}
\textcolor{black}{We have further extended our work to the three-user case and conducted experiments on it. In this case, each user group consists of three semantic users, namely user-N, user-M, and user-F. Similar to the two-user case, the information they require is superimposed and sent to each user through different links. In the simulation, we assume that user-N and user-M require CIFAR and MNIST images, respectively, while user-F requires textual data.}

\textcolor{black}{The end-to-end transmission results are given in Fig. \ref{three-user case}. It can be noted that all three users can achieve good reconstruction accuracy when the SNR is high. However, it is unsatisfactory in the low-SNR regime. Compared to the two-user case, more intra-group interference is introduced, thereby increasing the difficulty of detection. Therefore, improvements in model design and transmission mechanisms are worth investigating in future work to better serve larger user groups.}

\begin{figure*}[htbp]
	\centering
	\subfloat[Fitting result of $\boldsymbol{\xi}_{Cr}^{I}$]{
		\centering
		\includegraphics[height=1.5in,width=2in]{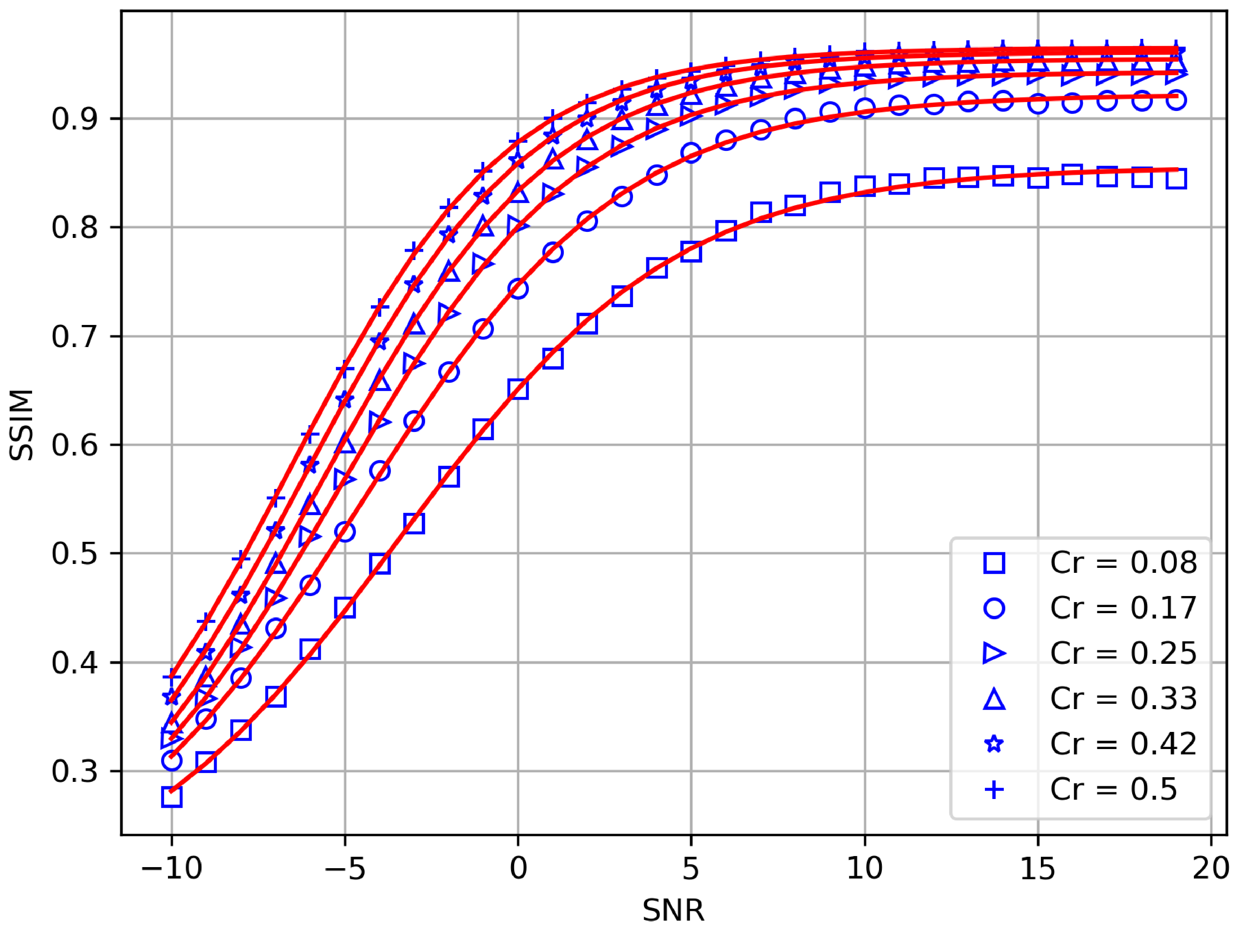}
		\label{ssimfitting}
	}
	\quad
	\subfloat[Fitting result of $\boldsymbol{\xi}_{K}^{S}$]{
		\centering
		\includegraphics[height=1.5in,width=2in]{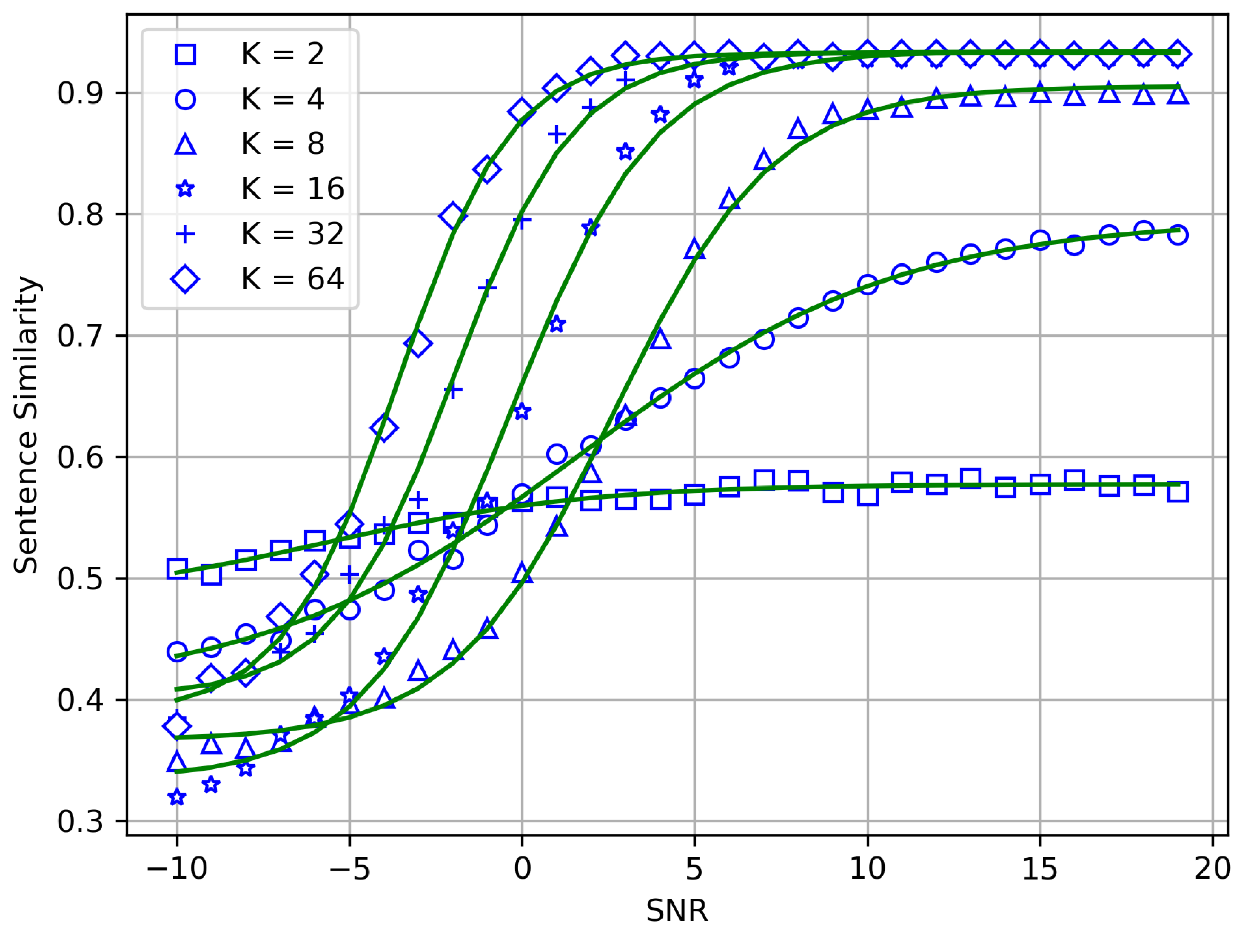}
		\label{sentencesimilarityfitting}
	}
        \\
 	\subfloat[Comparison of rate region between \textcolor{black}{the} proposed scheme and conventional scheme]{
		\centering
		\includegraphics[height=1.5in,width=2in]{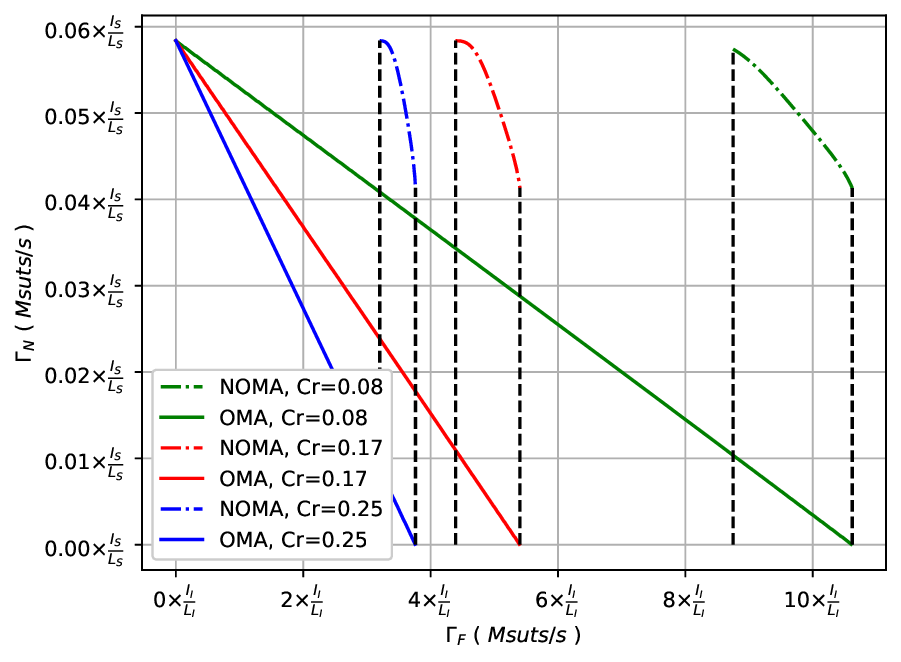}
		\label{rateregion}
	}
 	\quad
 	 	\subfloat[Comparison of power region between \textcolor{black}{the} proposed scheme and conventional scheme]{
		\centering
		\includegraphics[height=1.5in,width=2in]{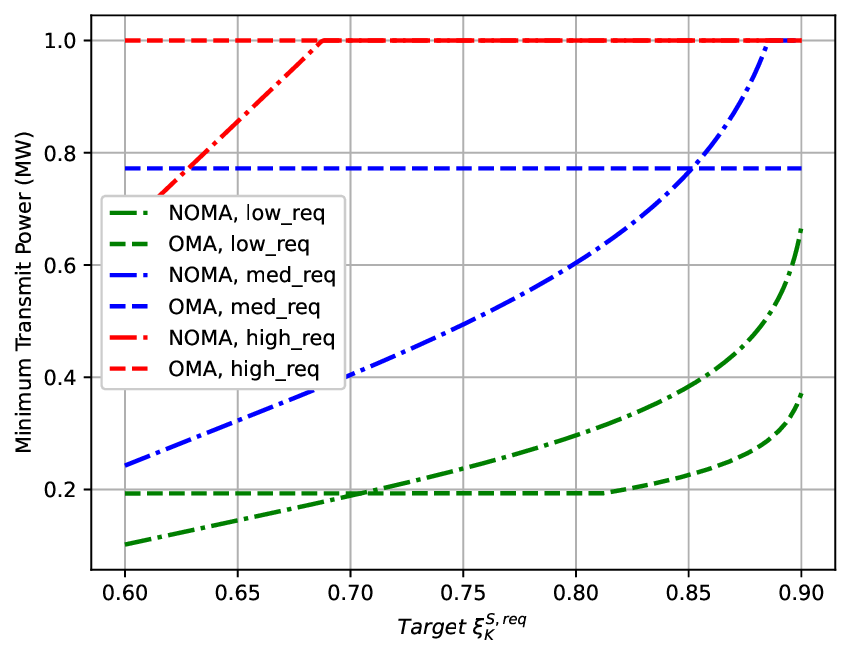}
		\label{powerregion}
	}
 	\centering
	\caption{The simulation result of rate region and power region, the function fitting result of $\boldsymbol{\xi}_{Cr}^{I}$ and $\boldsymbol{\xi}_{K}^{S}$ using logistic regression are given in (a) and (b). The rate region and power region of \textcolor{black}{the} NOMASC and the OMA scheme are given in (c) and (d).}
	\label{rateandpowerregion}
\end{figure*}
\subsection{Rate Region and Power Region}
For characterizing the rate region and power region of \textcolor{black}{the} NOMASC, an analytical form of performance metric sentence similarity and SSIM needs to be \textcolor{black}{determined}. Following the approach in \cite{seminoma}, a generalized logistic function is employed for fitting the curve:
\begin{align}\label{deqn_ex1a}
\overline{\boldsymbol{\xi}}_{K}(\gamma) = A_{K,1} + \frac{A_{K,2}-A_{K,1}}{1+e^{-\left(C_{K,1}\gamma+C_{K,2}\right)}}.
\end{align}
Using logistic regression, the coefficient in the function can be computed. We can get the approximate functions $\overline{\boldsymbol{\xi}}_{K}^{S}$ and $\overline{\boldsymbol{\xi}}_{Cr}^{I}$ for computing the semantic rate under different channel conditions. The function fitting results are depicted in Figs. \ref{ssimfitting} and \ref{sentencesimilarityfitting}, which show that for each $K$ and $Cr$, a group of coefficients can be derived \textcolor{black}{with good match with} the original data. The curves of \textcolor{black}{the} rate region and power region are plotted in Figs. \ref{rateregion} and \ref{powerregion} using the approximated functions $\overline{\boldsymbol{\xi}}_{K}^{S}$ and $\overline{\boldsymbol{\xi}}_{Cr}^{I}$. In the simulation, we set $\frac{P_{max}|\mathbf{h}_{N}|^{2}}{\sigma_{N}^{2}W} = 20\ dB$ and $\frac{P_{max}|\mathbf{h}_{F}|^{2}}{\sigma_{F}^{2}W}=16\ dB$, which makes the channel condition of user-N better than that of user-F. For characterizing the rate region, the accuracy requirement is set to $\boldsymbol{\xi}_{K}^{S,req}=0.6$ and $\boldsymbol{\xi}_{Cr}^{I,req}=0.7$. Using (\ref{24}) to (\ref{36}), the rate region can be computed and plotted in Fig. \ref{rateregion}. Under different $Cr$ settings, it \textcolor{black}{may be noted} that the feasible rate region of \textcolor{black}{the} NOMA always contains the corresponding rate region of OMA, which shows the advantage of the NOMA scheme and the sub-optimality of the OMA scheme. And the power region is calculated under the increase of sentence similarity requirements $\boldsymbol{\xi}_{K}^{S,req}$ and three cases with increasing S-Rate and SSIM requirements. In \textcolor{black}{the} 'low req' case, the requirements are set \textcolor{black}{as} $\boldsymbol{\xi}_{Cr}^{I,req}=0.65, \Gamma_{N}^{req}=0.063\times\frac{I_{S}}{L_{S}}\ (Msuts/s), \Gamma_{F}^{req}=3.13\times\frac{I_{I}}{L_{I}}\ (Msuts/s)$, in \textcolor{black}{the} 'med req' case, the requirements are set to $\boldsymbol{\xi}_{Cr}^{I,req}=0.68, \Gamma_{N}^{req}=0.069\times\frac{I_{S}}{L_{S}}\ (Msuts/s), \Gamma_{F}^{req}=4.13\times\frac{I_{I}}{L_{I}}\ (Msuts/s)$, and in \textcolor{black}{the} 'high req' case, the requirements are set to $\boldsymbol{\xi}_{Cr}^{I,req}=0.75, \Gamma_{N}^{req}=0.075\times\frac{I_{S}}{L_{S}}\ (Msuts/s), \Gamma_{F}^{req}=5\times\frac{I_{I}}{L_{I}}\ (Msuts/s)$. The resultant comparison results are shown in Fig. \ref{powerregion}. It can be \textcolor{black}{noted} that, \textcolor{black}{the} NOMA \textcolor{black}{has} gradually \textcolor{black}{shown} its advantage with the growth of rate and accuracy requirement. And it can also be observed that the OMA scheme is less sensitive to the increase of $\boldsymbol{\xi}_{Cr}^{I, req}$ since it only has to decrease the bandwidth allocated for user-F. While in the NOMA scheme, the whole bandwidth is shared by \textcolor{black}{the} two users, so when \textcolor{black}{the} required $\boldsymbol{\xi}_{Cr}^{I, req}$ grows, more power has to be allocated for user-F. Moreover, each OMA user can only be supplied \textcolor{black}{with} a portion of the bandwidth, which makes the power assumption for satisfying the rate \textcolor{black}{greater}.

\subsection{Complexity Analysis}
\begin{table*}[htbp]
\caption{Result of running time evaluation \label{tab:table2}}
\centering
\begin{tabular}{|c|c|c|c|c|c| }
\hline
Case & \multicolumn{5}{c|}{Scheme} \\
\hline
 & NOMASC & Conventional & Hybrid & JSCC-Q & DT-JSCC \\
\hline
Cifar\&Cifar & 0.51 ms & 142.30 ms & 18.69 ms & 0.73 ms & 1.32 ms\\
\hline
Cifar\&MNIST & 0.40 ms & 145.78 ms & 6.54 ms & 0.85 ms & 0.65 ms \\
\hline
Cifar\&Europarl & 0.44 ms & 122.53 ms & 33.06 ms & 0.65 ms & 0.84 ms \\
\hline
\end{tabular}
\end{table*}
\begin{figure}[htbp]
\centering
\includegraphics[width=2in]{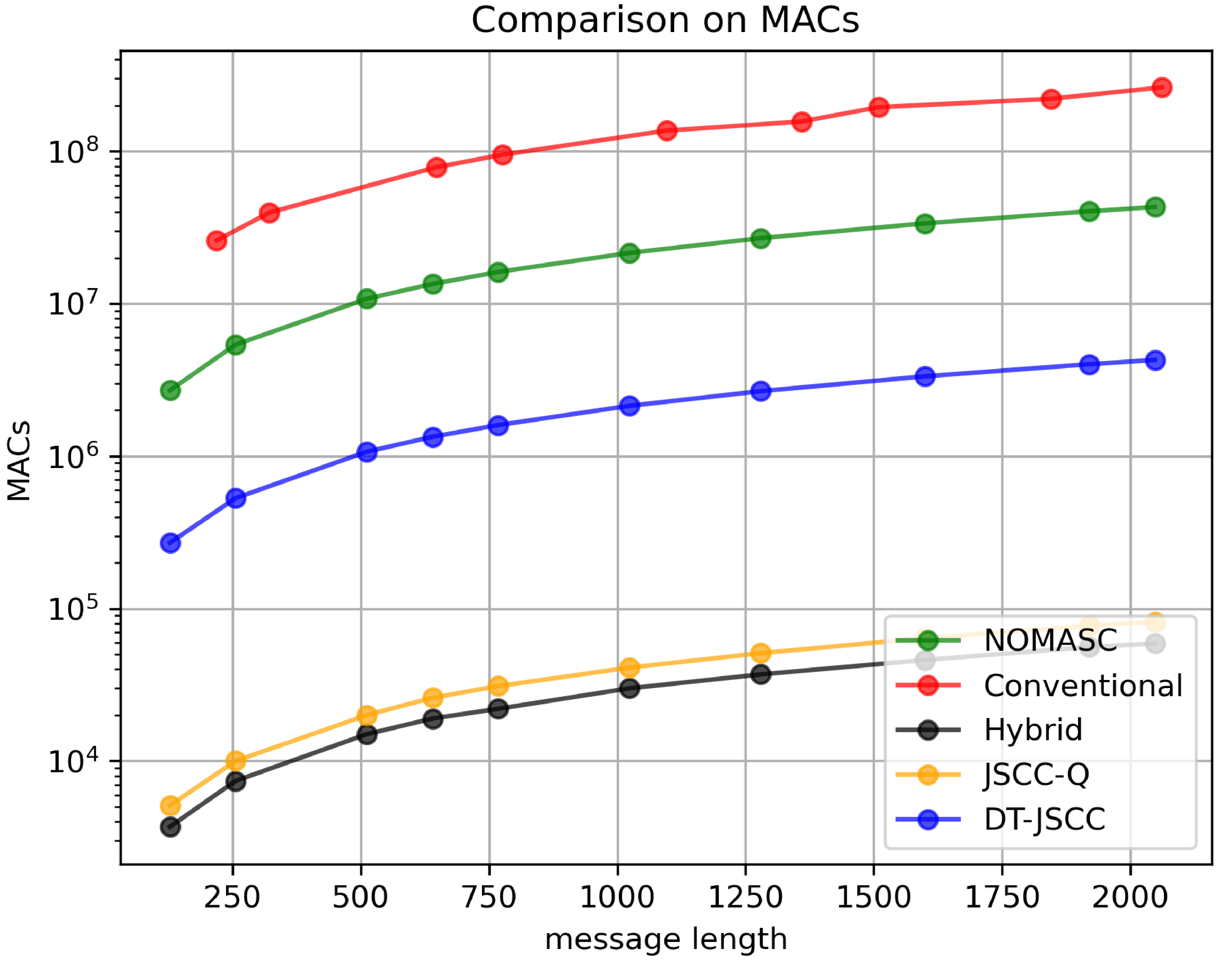}
\caption{Comparison \textcolor{black}{of} the complexity of various schemes in the MUD process. The complexity of the conventional method \textcolor{black}{is greater than} the NOMASC. And NOMASC yields the highest complexity among the learning-based methods. }
\label{MACcompare}
\end{figure}
Good-performing MUD algorithms may also have high time complexity \textcolor{black}{and} introduce a heavy burden for the hardware. Here, we \textcolor{black}{have firstly summarized} the running time of \textcolor{black}{various} methods for detecting one image or sentence in table \ref{tab:table2}. \textcolor{black}{As shown,} the detection time \textcolor{black}{for} NOMASC is the lowest and the conventional method \textcolor{black}{has taken} more time than the other learning-base methods. The computational complexity of all methods are also investigated in terms of the number of multiply-accumulate (MAC), which stands for one multiply and one accumulation operation. The computation method of MACs for LDPC and Turbo decoding is given in \cite{mac}. \textcolor{black}{As} shown in Fig. \ref{MACcompare}, the MACs for all methods \textcolor{black}{have shown} a linear increase along with the \textcolor{black}{length of} transmission message. Conventional methods seem to be inferior to the other methods in terms of complexity since LDPC code requires quite large computational resource. But the complexity of semantic compression and restoration can also be enormous when processing semantic source with larger size or using models with more complex structures.

\section{\textcolor{black}{Conclusion and Future Work}}
This paper \textcolor{black}{has introduced} a \textcolor{black}{novel} NOMA-based semantic communication system named NOMASC. The structure and detail design of \textcolor{black}{the} NOMASC is introduced. \textcolor{black}{In addition}, the proposed scheme's rate and power regions are analyzed. Extensive simulations have been conducted \textcolor{black}{to compare the} NOMASC \textcolor{black}{with} a series of methods under different performance metrics and channel settings. \textcolor{black}{Our} proposed system \textcolor{black}{has} proven to be able to serve \textcolor{black}{the} users with diverse types of datasets and holds strong robustness under \textcolor{black}{various testing} environments. The advantages of the proposed scheme \textcolor{black}{have also been illustrated} in terms of spectral and power efficiency. \textcolor{black}{Last but not the least}, the time complexity of the proposed scheme and each model in the system are \textcolor{black}{analyzed}. \textcolor{black}{Taken together}, \textcolor{black}{the} NOMASC \textcolor{black}{has enabled improvement of efficiency for} the non-orthogonal transmission of semantic information at an affordable complexity. 

\textcolor{black}{The system proposed in this paper currently only supports the fundamental case of pairing two or three users, and the types of transmitted sources have limited to text and images. Additionally, when the actual scenario differs significantly from the training setting, such as with significantly different datasets or large channel estimation errors, the system model may not be able to perform flexible transfer and its performance may decline.}

\textcolor{black}{Some future research directions may include optimizing the user pairing and quality-of-service (QoS)-aware resource allocation strategies, designing reconfigurable and adaptive model structure using meta-learning and developing novel multiple access strategies based on deep learning models. For example, learning to deeply fuse and superimpose semantic signals from multiple users, thereby further reducing transmission overhead.}

\begin{appendices}
\section{}\label{appendix}
Given that both $p_{z}$ and $x_{q}$ are integers, the minimum interval of the $x_{q}$ constellation set is 1. Therefore, to prove a zero point exists in $\mathcal{C}^{deq}$, we only have to prove that $-max(\mathbf{v}^{deq})\leq p_{z} \leq -min(\mathbf{v}^{deq})$, which is obtained using (\ref{xdeq}). Firstly, the range of $p_{z}$ can be given as follows:
\begin{equation}
\begin{aligned}\label{pzrange}
p_{z} &= round\left(-(s-d) * \frac{2^{m}-1}{2s}\right) \\
      &= round\left(- \frac{2^{m}-1}{2s/(s-d)}\right) \\
      &> - \frac{2^{m}-1}{2s/(s-d)} > - \frac{2^{m}-1}{2},\\
\end{aligned}
\end{equation}
where $2^{m}-1$ is always an odd number, and it is obvious that $p_{z}\leq0$ since $min(\mathbf{v}^{deq})\times f_{s}<0$. Using (\ref{pzrange})(\ref{xq}), the range of $min(\mathbf{v}^{deq})$ can be calculated as follows:
\begin{small}
\begin{equation}
\begin{aligned}\label{minxq}
min\left(\mathbf{v}^{deq}\right) &= clamp\left(round\left(- \frac{2^{m}-1}{2s/(s-d)}-p_{z}\right), 0, \ 2^{m} - 1\right) \\
      &\leq clamp\left(round(0, \ 0, \ 2^{m} - 1\right) = 0. \\
\end{aligned}
\end{equation}
\end{small}
In light of the fact that the mapping from $x$ to $x_{q}$ in (\ref{xq}) can be regarded as a non-decreasing function. And \textcolor{black}{by} using (\ref{clamp}), it can be seen that $min(\mathbf{v}^{deq})=0$, proving that $p_{z}\leq-min(\mathbf{v}^{deq})$. The range of $max(\mathbf{v}^{deq})$ can be computed as follows:
\begin{small}
\begin{align}
max(\mathbf{v}^{deq}) &= clamp\left(round\left((s+d)*\frac{2^{m}-1}{2s}\right), 0, 2^{m}-1\right)\notag\\
&=clamp\left(round\left(\frac{2^{m}-1}{2s/(s+d)}\right), 0, 2^{m}-1\right)\label{deqn_ex1a}\\
&>clamp\left(round\left(\frac{2^{m}-1}{2}\right), 0, 2^{m}-1\right)\notag\\
&>\frac{2^{m}-1}{2}.\notag
\end{align}
\end{small}
Thus, $-max(\mathbf{v}^{deq}) \leq p_{z}$ is proved using (\ref{pzrange}). Therefore, one constellation point $c_{i}$ can always be found in $\mathcal{C}^{deq}$ satisfying $c_{i}+p_{z}=0$.
\end{appendices}

\begin{IEEEbiography}[{\includegraphics[width=1in,height=1.25in,clip,keepaspectratio]{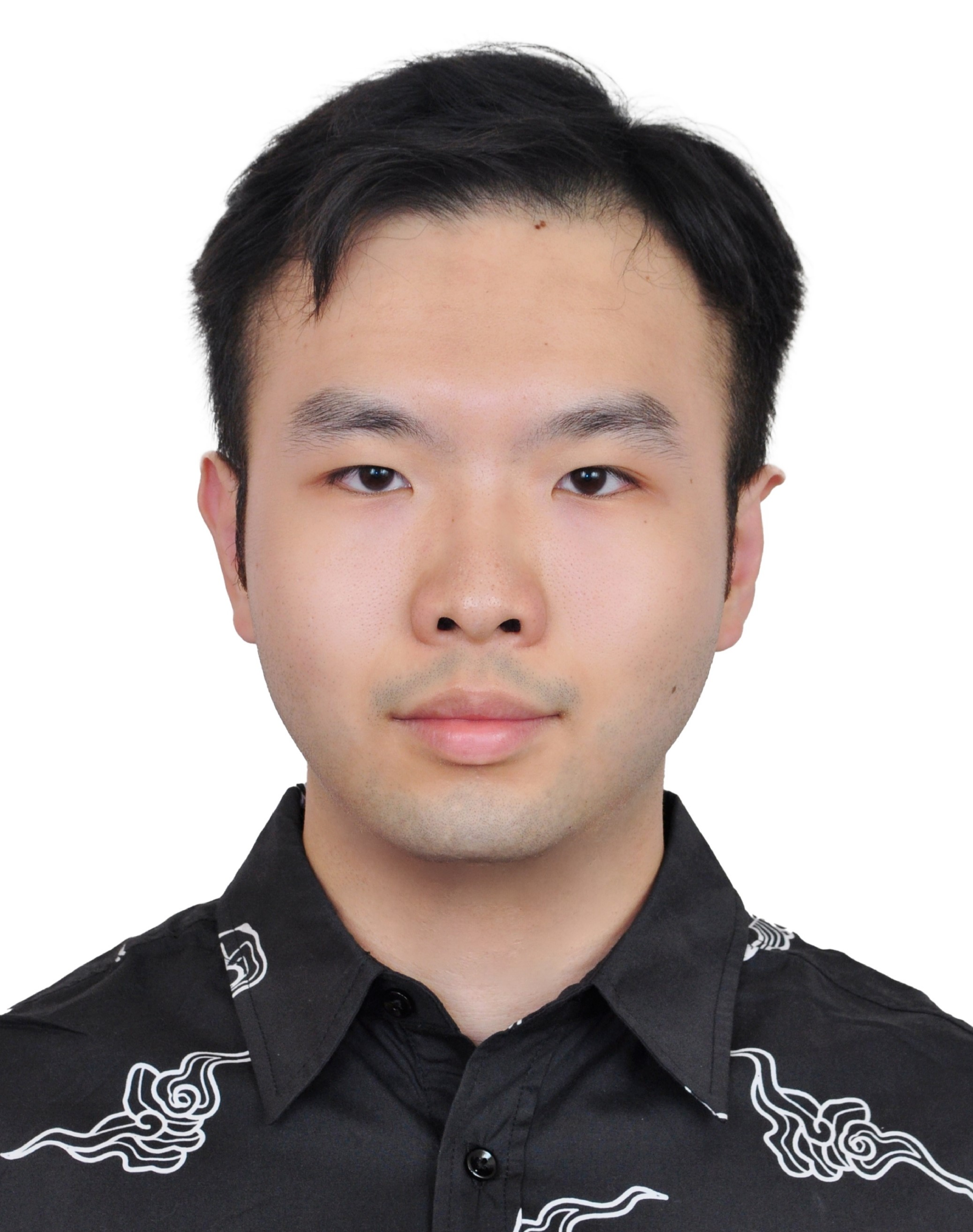}}]{Weizhi Li}
received the B.E. degree in Information Engineering from Beijing University of Posts and Telecommunications (BUPT), Beijing, China, in 2022, where he is currently pursuing the Ph.D. degree in Information and Communication Engineering. 

His research interests include next-generation networks, semantic communications and deep learning.
\end{IEEEbiography}

\begin{IEEEbiography}
[{\includegraphics[width=1in,height=1.25in,clip,keepaspectratio]{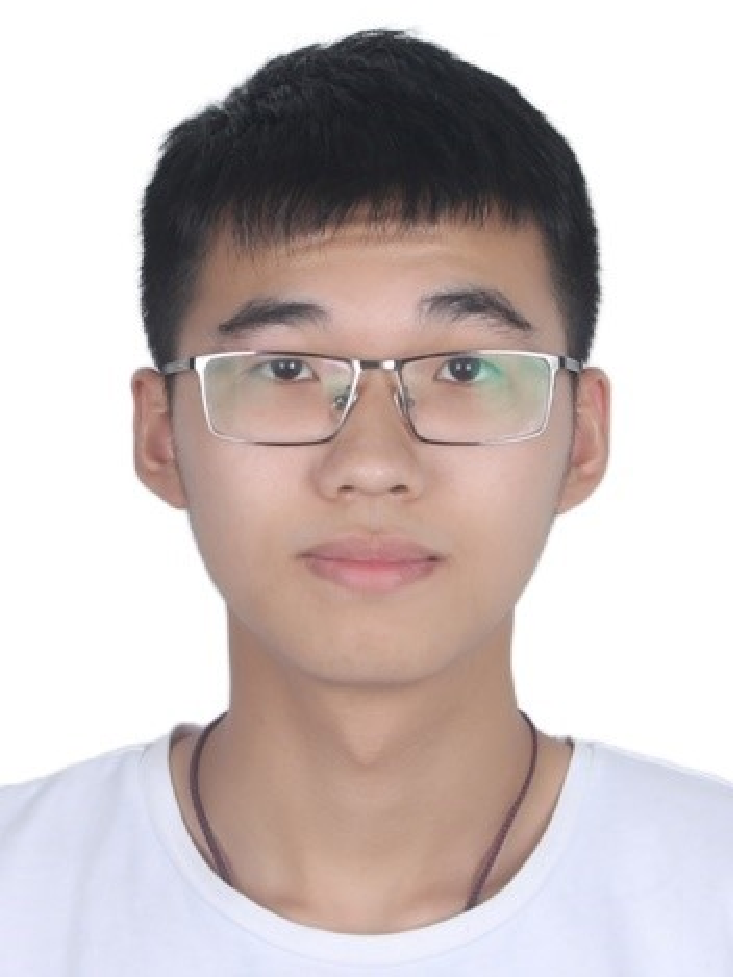}}]{Haotai Liang}
received the B.E. degree from the School of Electronic Information Engineering, Shenzhen University (SZU), Shenzhen, China, in June 2022.  He is currently pursuing the Ph.D. degree in Information and Telecommunications Engineering from Beijing University of Posts and Telecommunications (BUPT).
\end{IEEEbiography}

\begin{IEEEbiography}
[{\includegraphics[width=1in,height=1.25in,clip,keepaspectratio]{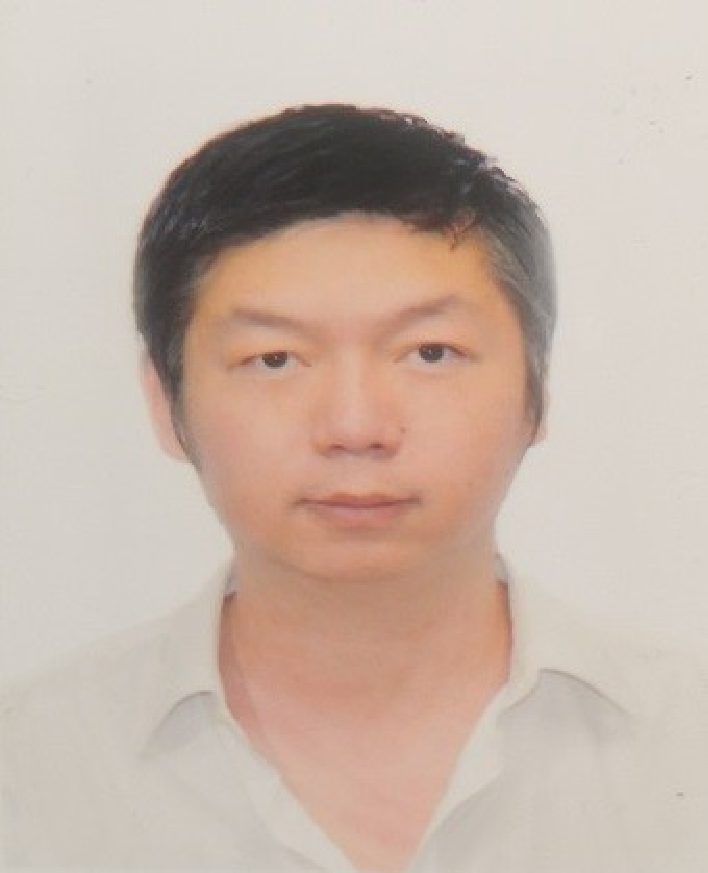}}]{Chen Dong}
received the B.S. degree in electronic information sciences and technology from the University of Science and Technology of China, Hefei, China, in 2004, the M.Eng. degree in pattern recognition and automatic equipment from the University of Chinese Academy of Sciences, Beijing, China, in 2007, and the Ph.D. degree from the University of Southampton, U.K., in 2014. After a post-doctoral researcher experience in Southampton, he used to work in Huawei Device Company Ltd., China. Since 2020, he has been working with the Beijing University of Posts and Telecommunications (BUPT). His research interests include applied mathematics, relay systems, channel modeling, and cross-layer optimization. He was a recipient of the Scholarship under the U.K.–China Scholarships for Excellence Programme and the Best Paper Award at the IEEE VTC 2014.
\end{IEEEbiography}

\begin{IEEEbiography}
[{\includegraphics[width=1in,height=1.25in,clip,keepaspectratio]{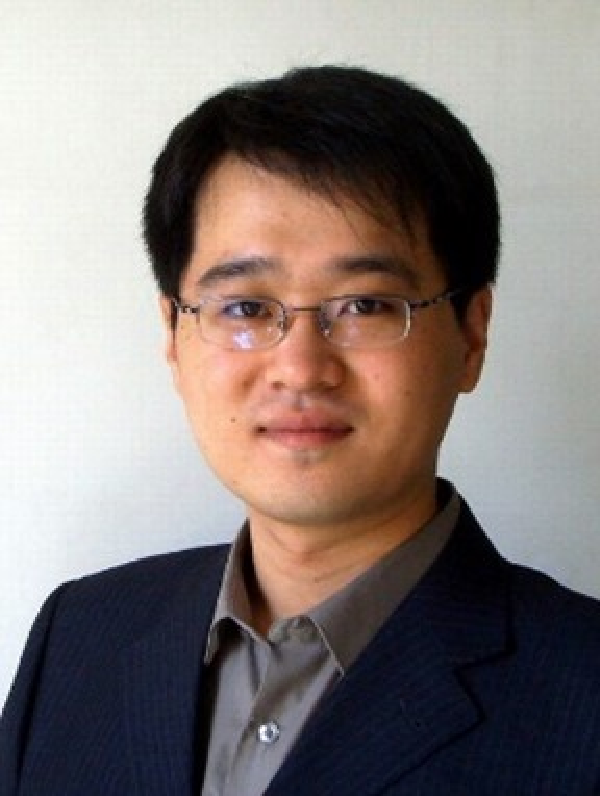}}]{Xiaodong Xu}
(Senior Member, IEEE) received his B.S degree in Information and Communication Engineering and Master’s Degree in Communication and Information System both from Shandong University in 2001 and 2004 separately. He received his Ph.D. degrees of Circuit and System in Beijing University of Posts and Telecommunications (BUPT) in 2007. He is currently a professor of BUPT, a research fellow of the Department of Broadband Communication of Peng Cheng Laboratory and a member of IMT-2030 (6G) Experts Panel. He has coauthored nine books/chapters and more than 120 journal and conference papers. He is also the inventor or co-inventor of 51 granted patents. His research interests cover semantic communications, intellicise communication system, moving networks, mobile edge computing and caching.
\end{IEEEbiography}

\begin{IEEEbiography}
[{\includegraphics[width=1in,height=1.25in,clip,keepaspectratio]{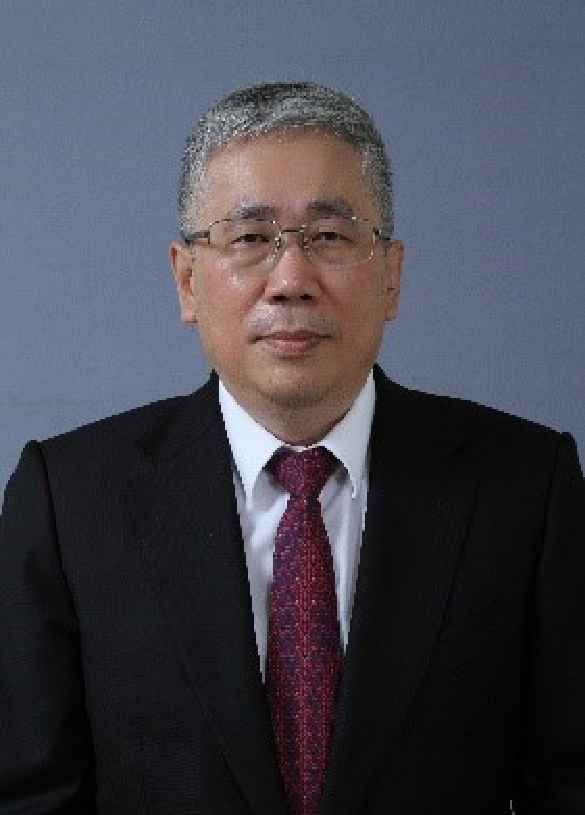}}]{Ping Zhang}
(Fellow, IEEE) received the Ph.D. degree in circuits and systems from the Beijing University of Posts and Telecommunications, Beijing, China, in 1990. He is currently a Professor with Beijing University of Posts and Telecommunications, the Director of the State Key Laboratory of Networking and Switching Technology, the Director of the Department of Broadband Communication of Peng Cheng Laboratory, a member of IMT-2020 (5G) Experts Panel, and a member of Experts Panel for China’s 6G development. He served as a Chief Scientist of National Basic Research Program (973 Program), an expert in Information Technology Division of National High-Tech R\&D Program (863 Program), and a member of Consultant Committee on International Cooperation of National Natural Science Foundation of China. His research interests mainly focus on wireless communication. Prof. Zhang is an Academician of the Chinese Academy of Engineering.
\end{IEEEbiography}
\begin{IEEEbiography}
[{\includegraphics[width=1in,height=1.25in,clip,keepaspectratio]{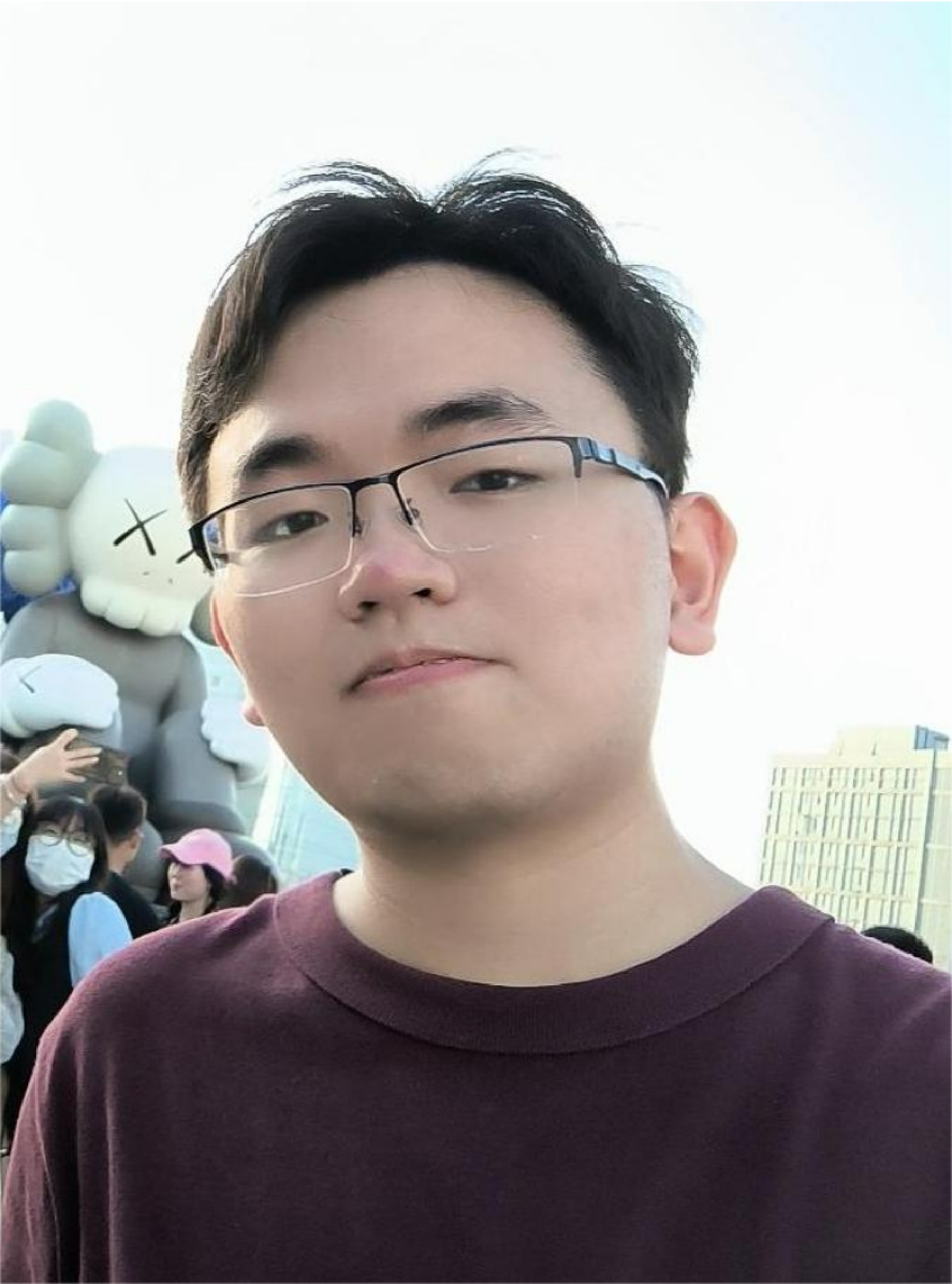}}]{Kaijun Liu}
received the B.E. degree from the School of Science, Beijing University of Posts and Telecommunication (BUPT), Beijing, China, in June 2021. He is currently pursuing the M.E.
degree in the School of Information and Communication Engineering in Beijing University of Posts and Telecommunication.
\end{IEEEbiography}

\vfill

\end{document}